\documentclass[useAMS,usenatbib]{mn2e}

\usepackage[english]{babel}
\usepackage{graphicx}
\usepackage{amssymb}
\bibpunct{(}{)}{;}{a}{}{,}
\def\grs{GRS~1915+105}

\title[\grs: A comparison of the plateau state to canonical hard
states]
{%High luminosity jets: Can the \grs\ plateau state be compared to the canonical hard state?
Is the plateau state in \grs\ equivalent to canonical hard states?
}

\begin{document}
\author[P.C.N. van Oers et al.]{Pieter van Oers$^{1,2}$, Sera Markoff$^1$, Dipankar Maitra$^{1,3}$, Farid Rahoui$^{4,5,6}$, \and Michael Nowak$^7$, J\"{o}rn Wilms$^8$,  Alberto J. Castro-Tirado$^9$, Jerome Rodriguez$^4$,  \and Vivek Dhawan$^{10}$, and Emilios Harlaftis$^{11}$\thanks{In Memoriam}\\
$^1$Astronomical institute``Anton Pannekoek", University of Amsterdam,
Science Park 904, 1098 XH, The Netherlands\\
$^2$New affiliation: School of Physics \& Astronomy, University of Southampton, Highfield, Southampton SO17 1BJ, United Kingdom\\
$^3$New affiliation: Department of Astronomy, University of Michigan, 500 Church St., Ann Arbor, MI 48109, USA\\
$^4$Laboratoire AIM,CEA/IRFU, Universit\'{e} Paris Diderot, CNRS/INSU, CEA Saclay, DSM/IRFU/SAp, F-91191 Gif sur Yvette, France\\
$^5$AstroParticule \& Cosmologie (APC) / Universit\'{e} Paris VII / CNRS / CEA / Observatoire de Paris, B\^{a}t. Condorcet,\\ \hfill 10 rue Alice Domon et L\'{e}onie Duquet, 75205 Paris Cedex 13, France\\
$^6$Department of Astronomy \& Harvard-Smithonian Center for Astrophysics, Harvard University, 60 Garden street, Cambridge, MA 02138, USA\\
$^7$Kavli Institute for Astrophysics and Space Research, Massachusetts Institute of Technology, Cambridge, USA\\
$^8$Dr. Karl Remeis-Sternwarte, Astronomisches Institut, Universit\"{a}t Erlangen-Nuremberg, Sternwartstr. 7, 96049, Germany\\
$^9$Instituto de Astrof\'{\i}sica de Andaluc\'{\i}a (IAA-CSIC), Glorieta de la Astronom\'{\i}a s/n, 18008 Granada, Spain\\
$^{10}$National Radio Astronomy Observatory, 10003 Lopezville Road, Socorrow, NM 87801, USA\\
$^{11}$Institute of Space Applications and Remote Sensing, National Observatory of Athens, PO Box 20048, Athens 11810, Greece}

%\pagerange{\pageref{firstpage}--\pageref{lastpage}} \pubyear{2008}

\maketitle

\label{firstpage}

\begin{abstract}\grs\ is a very peculiar black hole binary that exhibits accretion-related states that are not observed in any other stellar-mass black hole system.  One of these states, however -- referred to as the plateau state -- may be related to the canonical hard state of black hole X-ray binaries. Both the plateau and hard state are associated with steady, relatively lower X-ray emission and flat/inverted radio emission, that is sometimes resolved into compact, self-absorbed jets. However, while generally black hole binaries quench their jets when the luminosity becomes too high, \grs\ seems to sustain them despite the fact that it accretes at near- or super-Eddington rates. In order to investigate the relationship between the plateau and the hard state, we fit two multi-wavelength observations using a steady-state outflow-dominated model, developed for hard state black hole binaries. The data sets consist of quasi-simultaneous observations in radio, near-infrared and X-ray bands. Interestingly, we find both significant differences between the two plateau states, as well as between the best-fit model parameters and those representative of the hard state. %In particular we find that the plateau state jets must be not only extremely luminous, but also extremely magnetically dominated. 
We discuss our interpretation of these results, and the possible implications for \grs's relationship to canonical black hole candidates.
\end{abstract}


\section{Introduction}
\label{sec:grsintro}

The microquasar \grs\ was discovered on 15 August 1992, by the
WATCH all-sky monitor, aboard the Russian GRANAT satellite
(\citealt*{1992IAUC.5590....2C}; \citealt{CastroTiradoetal1994}).  It is a hard X-ray transient located in
the constellation of Aquila, at $l=45.37^{\circ}$, $b=-0.22^{\circ}$,
and was the first stellar mass accreting black hole binary (BHB)
discovered to
exhibit superluminal velocities in its radio emitting-ejecta
\citep{1994Natur.371...46M}.  The obvious parallels to the jets in Active
Galactic Nuclei (AGN) led to this source being classified as a
``microquasar'' \citep{1998Natur.392..673M}.     Subsequent observations with instruments onboard the
\textit{Rossi X-ray Timing Explorer} (\textit{RXTE}) have revealed a richness in
variability, distinguishing \grs\ from every other known BHB,
over which astronomers are still puzzling to this day.

The longer term X-ray variability exhibits ``dipping'' that is thought to be
associated with the recession and regeneration of the inner parts of the accretion disc, perhaps caused by the onset of
thermal-viscous instabilities \citep{Bellonietal1997a,Bellonietal1997b}. Other
models have interpreted the spectral changes as resulting from the
disappearance of the corona \citep{1998PhDT........33C,2008ApJ...675.1436R,2008ApJ...675.1449R}, or from the
dissipation of magnetic energy via magneto-hydrodynamical plasma
processes \citep[e.g.][]{2004ApJ...607..410T}.  

Beyond the dipping behaviour, \citet{Bellonietal2000} were able to
classify all variability patterns stretching over more than a year
into only twelve classes (\citealt{2002MNRAS.331..745K, Hannikainenetal2005} later
identified two more classes), based on colour-colour diagrams and
light curves. Ten of these twelve classes can be understood as the
interplay of two or three of three basic states, designated as states
A, B and C. The remaining two classes, $\phi$ and $\chi$, do not show
state transitions and appear exclusively within states A and C
respectively.  State A displays the highest flux and the softest spectrum while state C displays the lowest flux and the hardest spectrum. Although state B never lasts for more than a few hundred seconds, $\phi$ and $\chi$ episodes can persist for days or short intervals of $<$ 100 s. 

Aside from the existence of so many distinct accretion states,
\grs\ appears similar to other BHBs.  Thus there has been much
discussion \citep*[e.g.][]{ReigBellonivanderKlis2003} about the extent to which any of the states found in \grs\ correspond
to the ``canonical" states (see e.g. \citealt{HomanBelloni2005,RemillardMcClintock2006}
for definitions) found in the average BHB. BHBs generally spend the majority of their time in
the Hard State (HS), which is associated with a low accretion rate,  a hard X-ray power law component with 
photon index 1.4$\le\Gamma\le$2.1, and steady radio emission with an
optically thick, flat-to-inverted spectrum.  Sources that persist in the HS
for several weeks generally show 
clear correlations between the
radio and X-ray luminosities ($L_R\propto L_X^{0.7}$)
(\citealt{2000arxt.confE.106C}; \citealt{Corbeletal2003};
\citealt*{GalloFenderPooley2003}). More recently, evidence for a similar
correlation between the near-infrared (NIR) and X-ray bands has 
also been found
\citep{Russelletal2006,Coriatetal2009}. %see page 32 of \cite{2003astro.ph..6213M}

Similar to the other BHBs, \grs\ also displays periods of relatively
hard, steady X-ray emission, but in contrast to the HS, only for about half the observation
time \citep{2001ApJ...558..276T}.
\citet{Bellonietal2000} identify these intervals of decreased
variability with long C state episodes. The subset of state C/ class
$\chi$ observations that are radio-bright are referred to
as the \emph{plateau} state (ibid.), but elsewhere have variously been
been referred to as the radio-loud, radio-plateau or radio loud
low/hard X-ray state \citep{2001ApJ...556..515M}, the type II hard
steady state \citep{2001ApJ...558..276T} and $\chi_{\rm RL}$
\citep{2000A&A...362..691N}.

The plateau state was first described by \citet{1996ApJ...467L..81F}
and its description was later refined in
\citet{Fenderetal1999}.  The state can assert itself in a period as
short as a day and is characterised by a decrease in X-ray flux
density and an increase in radio flux density to a steady value, 
typically $\sim10-100$ mJy. Ample evidence supports the fact that the
plateau state X-ray and radio luminosities are indeed correlated, with
an increasing delay from X-ray to infrared (IR) to radio emission
\citep[e.g.][]{2002MNRAS.331..745K}.
As in the HS the plateau radio spectrum is optically thick, and
AU-scale self-absorbed compact jets have been spatially resolved using
Very Long Baseline Interferometry (VLBI) (\citealt*{2000ApJ...543..373D}; \citealt{Fuchsetal2003b}). Another marker of the plateau state
comes from timing analyses, where 1-10 Hz %type-C 
quasi-periodic
oscillations (QPOs) are present \citep{2000A&A...360L..25R}. The exact
frequency of these QPOs appears to be inversely correlated with the
radio and soft X-ray flux \citep{2003A&A...397..711R,Rodriguezetal2002}.

In line with these arguments, it is tempting to identify the plateau
state as \grs's analog of the HS.  However, although the
plateau state and the HS share many similarities, it does display some
distinct properties that cannot be ignored.  For instance, while the
BHBs in the HS usually have a luminosity of $\lesssim10\%$ $L_{\rm
  Edd}$, the average luminosity observed in the plateau state is $\sim
L_{\rm Edd}$.  Moreover, the plateau X-ray photon index is never
very hard, with $\Gamma\sim$ $1.8-2.5$
\citep{2004ARA&A..42..317F}. Finally, \citet*{ReigBellonivanderKlis2003} argue that
the canonical HS is never seen in \grs\ because a power-law
tail (with no exponential cut off) is always present in the plateau state hard X-rays. Although the origin of such
tails in the softer states of canonical BHBs is still under debate, it is generally not associated with the HS.

In this work, we seek to make a more quantitative comparison between
the plateau state found in \grs\ and the HS found in more
typical black holes, in the context of an outflow-dominated model that
has successfully described broadband data from several BHBs. In Section \ref{sec:grsobs} we describe the multi-wavelength data and the reduction methods. We discuss what physical parameters we use for modelling and why in Section \ref{sec:grsparams} and the model itself, together with the results in Section \ref{sec:grsmodres}. We put our work into context with the findings of others in Section \ref{sec:grsdisc}, before drawing our final conclusions in Section \ref{sec:grsconcl}.

\begin{figure}
\includegraphics[width=84mm]{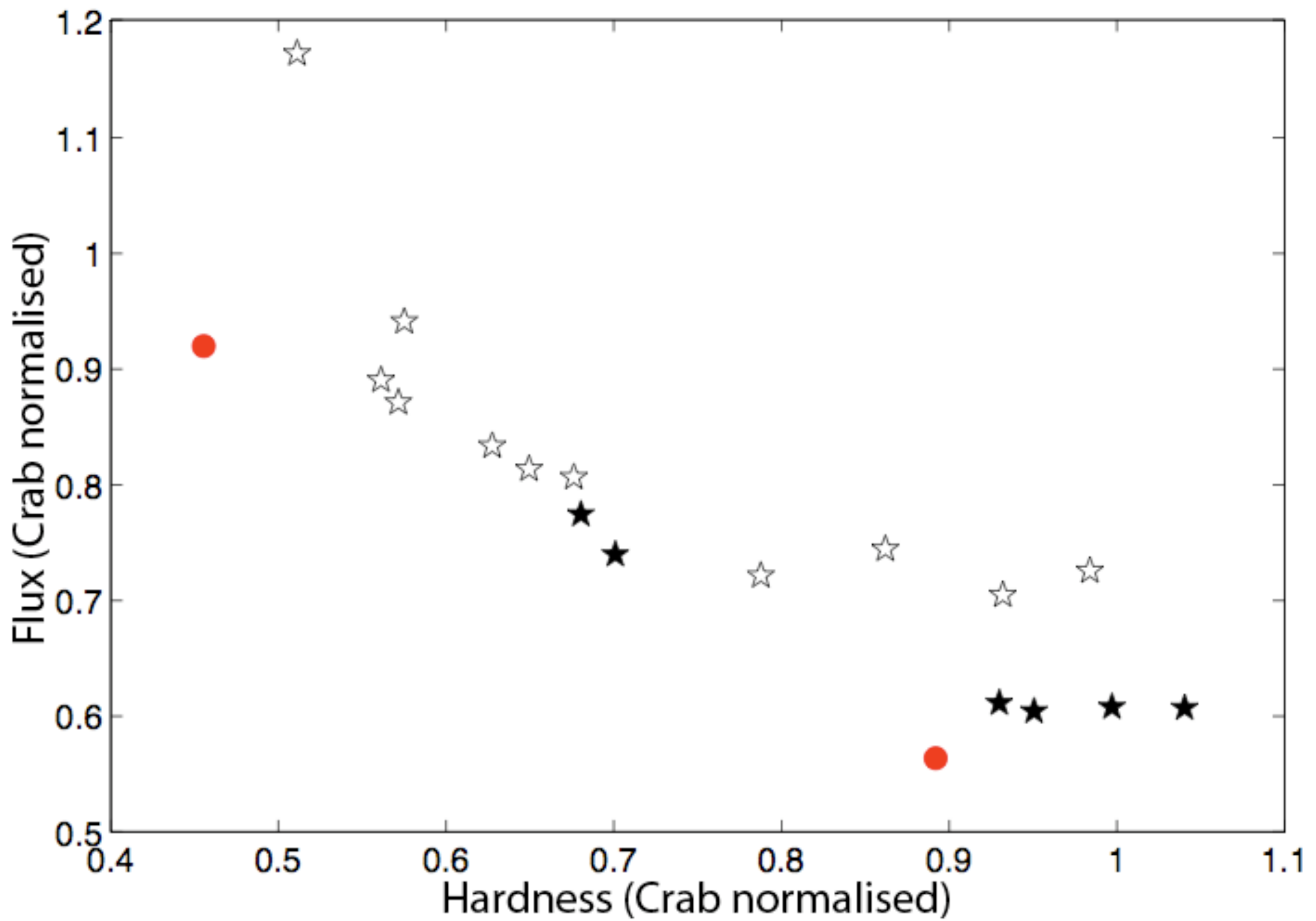}
 \caption{HID comparing our observations (filled circles) to the plateau states from \citet{Bellonietal2000} (data courtesy of T. Belloni; open stars are $\chi1$, closed stars are $\chi3$). The soft, higher-luminosity red dot is ObsID 90105-07-02-00, the lower is ObsID 40403-01-09-00.  The plot shows the hardness and flux, normalised to Crab units, calculated by taking the ratio and the sum of the \textit{RXTE} PCU2 rates in two bands: channel 0-35 ($\sim2-15$ keV) and 36-49 ($\sim16-21$ keV). Each band is divided by the Crab rate in the same band, on the same day, simulating a photon index of 2.1 and utilising a normalisation of 10 photons cm$^{-2}$ s$^{-1}$ keV$^{-1}$ at 1 keV.}
 \label{fig:HID}
\end{figure}

\section{Observations and Data Reduction}
\label{sec:grsobs}

\begin{table*}
\begin{minipage}{162mm}
\caption{Observations included in dataset 1 and 2, with \textit{RXTE} ObsID 40403-01-09-00 and 90105-07-02-00 respectively. The $H$ and $K$ bands, obtained using the Cooled Grating Spectrometer at UKIRT, are at 1.455-2.094 and 1.906-2.547 $\mu$m, respectively ($J$ band measurements were also done, but these are unusable due to high interstellar extinction). The $K$ band observation in dataset 2 was done (more than half a day before ObsID 90105-07-02-00) with the ANDICAM at CTIO. The GBI measurements are 2.25 \& 8.3 GHz and the flux densities used are 30 sec vector scan averages. The Ryle telescope operates at 15 GHz and the flux density used is the average of $\sim$13 ks worth of five minute integrations.}
\label{tab:grsobs}
\begin{tabular}{@{}lclccccclccc}
\cline{3-7}\cline{9-12}
		&&\multicolumn{5}{c}{{\bf Dataset 1} 1999 July 8 / MJD 51367}&\hspace{.3cm}	& \multicolumn{4}{c}{{\bf Dataset 2} 2005 April 13 / MJD 53473}\\
		\cline{3-7}\cline{9-12}
\hline
Band	&& \multicolumn{2}{c}{Instrument}& MJD start	& UTC		& Duration&	& Instrument 		& 	MJD start	& Time		& Duration\\
\hline
X-ray	&& \textit{RXTE}&		& 51367.2737	& 06:34:13	& 14435 sec	&& \textit{RXTE}			& 53473.054		& 01:09:05	& 6769 sec\\
	%	&& &($J$)			& 51367.5753	& 13:48:27	& 64 min		&& 	& 	& 	&\\
NIR		&& UKIRT	&($H$)	& 51367.4821	& 11:34:10	& 80 min		&&	CTIO ($K$)		&53472.3216	&07:43:06	& 25 min\\
		&&	&($K$)		& 51367.8216	& 07:56:59	& 64 min		&&&&&\\
Radio	&\hspace{.6cm}	& GBI&		& 51367.371	& 09:02:53	& 30 sec		&& Ryle 			& 53473.098	& 02:21:07 & 219 min\\
\hline
\end{tabular}
\end{minipage}
\end{table*}

\begin{figure}
%\begin{minipage}{168mm}
\includegraphics[width=84mm]{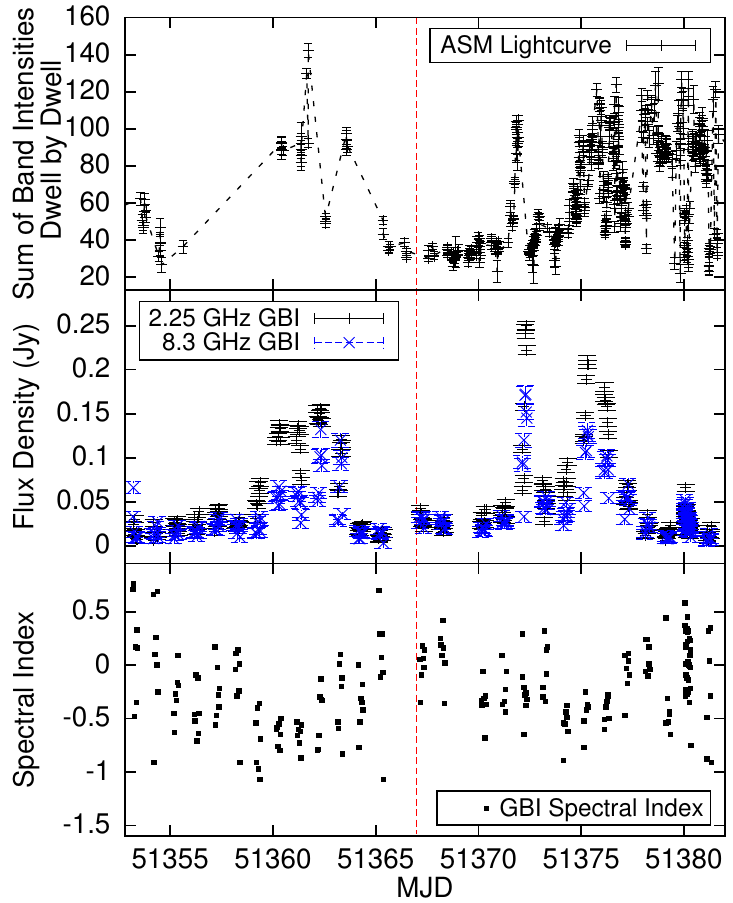}
 \caption{\grs\ X-ray and radio lightcurves for the 1999 data set. Also the radio spectral index, as measured by the Green Bank Interferometer is shown. The red dashed line indicates the time of the \textit{RXTE} observation.}
 \label{fig:datasetA}
\end{figure}
\begin{figure}
\includegraphics[width=84mm]{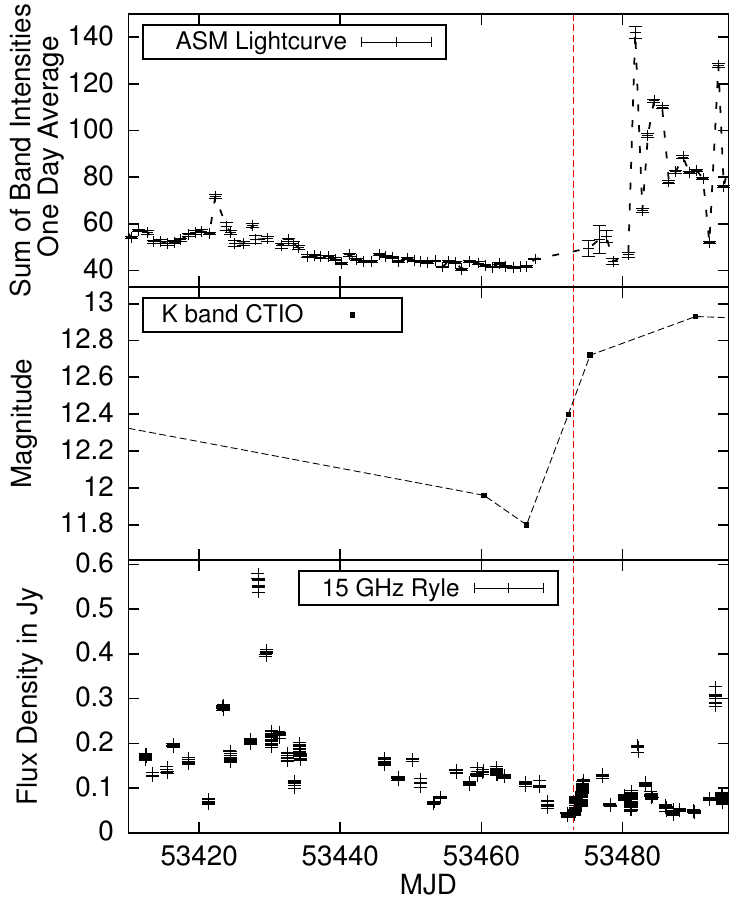}
 \caption{\grs\ X-ray and radio lightcurves for the 2005 data set. The radio lightcurve was measured by the Ryle Telescope \citep{PooleyFender1997}. The red dashed line indicates the time of the \textit{RXTE} observation.}
 \label{fig:datasetB}
% \end{minipage}
\end{figure}

\begin{figure}
\includegraphics[width=80mm]{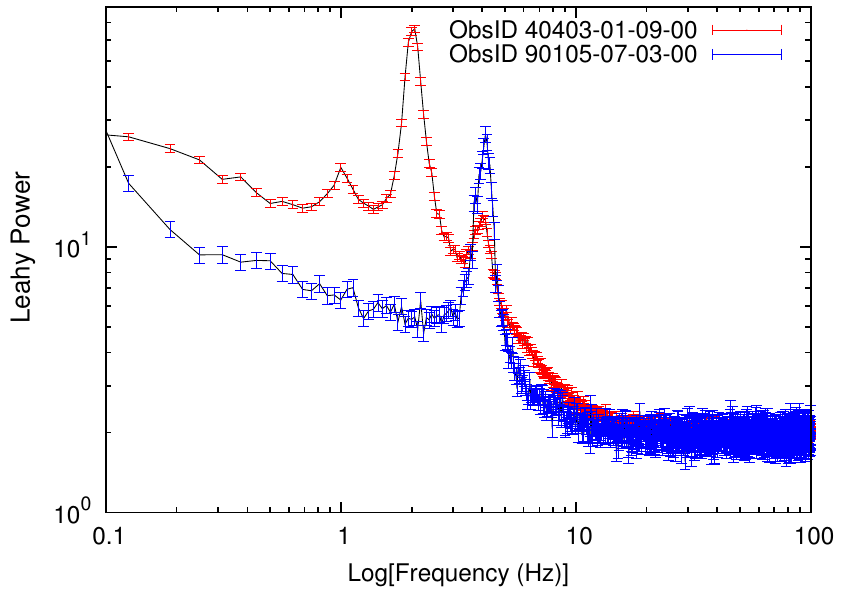}
 \caption{Normalised \citep[following][]{Leahyetal1983} power density spectra  showing the type-C QPOs in the X-rays for both datasets, indicating the observations are plateau state. (Courtesy of P. Soleri)}
 \label{fig:qpos}
\end{figure}

Based on the lightcurves and its position in the Hardness Intensity
Diagram (HID) diagram (see Figures \ref{fig:HID}, \ref{fig:datasetA} \&
\ref{fig:datasetB}), \grs\ was in a class $\chi$ state on July
8th 1999 (dataset 1) and April 13th 2005 (dataset 2). The presence of
a 2-5 Hz QPO on both occasions (see Figure \ref{fig:qpos}) corroborates this fact.  During the 1999 episode, quasi-simultaneous radio and near-infrared observations were
performed using the United Kingdom Infra-Red Telescope (UKIRT) and the Green Bank Interferometer (GBI). The 2005 dataset also holds observations of the Cerro Tololo Inter-American Observatory (CTIO) and Ryle telescope. Details on the observations and reduction methods are found in the individual Sections
and Table \ref{tab:grsobs}. The 2005 \textit{RXTE} data is also extensively analysed and discussed in \citet{2008ApJ...675.1436R,2008ApJ...675.1449R}

\subsection{X-ray: \textit{RXTE} data reduction}

We use data from two instruments on board the \textit{RXTE}: the Proportional Counter Array (PCA; \citealt{Jahodaetal2006}) and the High Energy X-ray Timing Experiment (HEXTE; \citealt{1998ApJ...496..538R}). The
X-ray data have been reduced using \texttt{HEASOFT}, version 6.3.1.,
applying standard extraction criteria: a pointing offset of $<
0.01^{\circ}$ from the nominal source position, and a source elevation
of $> 10^\circ$. The exclusion time for the South Atlantic Anomaly
(SAA), however, was only 10\,min, as \grs\ is a very bright
source. For the same reason we disregarded the top layer and imposed a
maximum ``electron ratio'' of 2, to take into account contamination by
source photons. All HEXTE data products were dead-time
corrected. Correction for the PCA dead-time was also carried out; PCU 4
was inactive for half of the 1999 observation, while during the other
half, both PCU 1 and 4 were inactive. The 2005 observation was done
with PCU 0 and 2 only. Due to uncertainties in the first four PCA
channels, we only include PCA data above 3\,keV. The PCA calibration falls off rapidly above $\sim25$ keV and since HEXTE provides reliable data for energies $\ga 20$\,keV we ignored the PCA data above
22\,keV. For the PCA, standard2f mode data were used. The PCA
background was modelled using the
\texttt{pca\_bkgd\_cmbrightvle\_eMv20051128} model. Only HEXTE data
above 20\,keV were used due to the uncertainty of the response matrix
below these energies.  At high energies, the spectrum was ignored
above 200\,keV. We rebinned the X-ray data to a minimum S/N of 4.5.

\subsection{NIR: UKIRT data reduction and CTIO}

 On 1999 July 08, medium-resolution spectroscopic observations of  
\grs\ were performed
 with the Cooled Grating Spectrometer (CGS4) mounted on the United  
Kingdom
InfraRed Telescope (UKIRT, P.I. Harlaftis). Twenty-four spectra in the  
B1 and eighteen in the B2 filters were acquired,
each time at two nodding positions, giving a total exposure time of  
1260~s in B1 and 1080~s in B2.
Moreover, the telluric standard HD~176232, a F0p main sequence star,  
was observed in the same conditions.

All the spectra were reduced with \textit{IRAF} through bad pixel  
correction, flat fielding, and sky subtraction.
The target spectra were then extracted and wavelength calibrated by  
extracting, in the same condition,
krypton and argon spectra in B1 and B2, respectively. They were  
combined and telluric feature-corrected
through the division by the standard star spectrum. We finally  
multiplied the corrected spectrum by a F0V star one
from the ISAAC synthetic spectra library\footnote{\texttt{http://www.eso.org/sci/facilities/paranal/instruments/isaac/tools/lib/index.html 
}},
resampled to the CGS4 resolution and flux-scaled to match the  
HD~176232 magnitudes in H and Ks.
The uncertainty of the flux calibration obtained this way is about 5\%.
 
 \begin{figure}
\includegraphics[width=84mm]{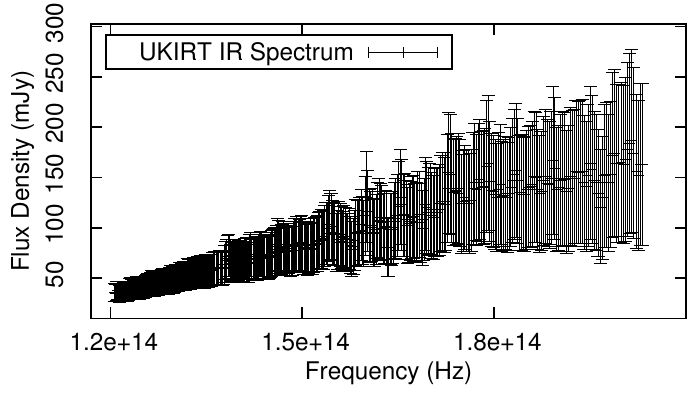}
 \caption{NIR spectrum of the 1999 data set}
 %, data only (top) and best fit model (bottom). The dashed and dotted lines in the bottom plot are the contribution of the post-shock synchrotron and the companion blackbody respectively, while the total model flux is represented by the solid line.
 \label{fig:ir}
\end{figure}

The spectra were then dereddened using the extinction laws given in \citep*{1989ApJ...345..245C} and \citep{ChiarTielens2006}, assuming a hydrogen column density of $N_{\rm
  H}$=3.5$\times10^{22}$ cm$^{-2}$, i.e. Av=19.6+/-1.7  (\citealt{2004A&A...414..659C}. The errors resulting from using the two different methods were then added in quadrature. The overall propagated uncertainties, combining flux calibration and dereddening, represent about 20\% of the flux. The final spectrum is shown in Figure \ref{fig:ir}.
 
The UKIRT spectrum included in dataset 1 was also re-binned, %(see Figure\ref{fig:ir})
 because the spectral resolution is well beyond what we
need for continuum fitting (see Section~\ref{sec:grsmodres}). 

The 12.40 mag K band point in dataset 2 was taken using the CTIO 1.3m
telescope using the ANDICAM (A Novel Double-Imaging CAMera) detector \citep*{2007ApJ...657..409N}. 
It was dereddened using the same method as the UKIRT data. However we chose the error bars to account for the fact that \grs\ may have been approaching a state transition: The source magnitude  
magnitude varies from 11.8 to 12.72 in about 9 days (see Figure \ref{fig:datasetB}), suggesting a dereddened flux density of 67.85$\pm27.25$ mJy. On MJD 53473 we also see the flux in radio increasing (see below).

\subsection{Radio: GBI and Ryle Telescope}

Dataset 1 includes two GBI data points and dataset 2 includes a data
point from Ryle Telescope. Lightcurves from these telescopes are shown
in Figures \ref{fig:datasetA} and \ref{fig:datasetB}, respectively.

For the GBI data, a flux density calibration procedure similar to that reported in Waltman et al (1994) has been employed here. The flux densities of 0237-233, 1245-197, and 1328+254 were determined using observations of 1328+307 (3C 286). The flux density of 3C 286 was based on the scale of \citet{Baarsetal1977}, and the assumed values were 11.85 Jy at 2.25 GHz and 5.27 Jy at 8.3 GHz\footnote{\texttt{http://www.gb.nrao.edu/fgdocs/gbi/plgbi/README}}. On MJD 51367.371 the GBI flux densities were 25$\pm$4 mJy at 2.25 GHz and 30$\pm$6 mJy at 8.3 GHz. Similar behaviour was shown during a plateau state of April 2000 \citep{2002ApJ...571..918U}.

The Ryle Telescope operates at 15 GHz and it observed \grs\ (simultaneously with ObsID 90105-07-02-00) from MJ D 53473.098 to MJD 53473.417. The data used here are from \citet{2008ApJ...675.1436R} and were reduced following \cite{PooleyFender1997}; observations of Stokes I+Q were interwoven with those of a nearby phase calibrator (B1920+154). The flux-density scale was set by reference to 3C48 and 3C286, and should be consistent with that defined by \citet{Baarsetal1977}.

On MJD 53473 we see the average radio flux density increasing from $44.9\pm3.0$ to $70.4\pm4.3$ mJy (\citealt{2008ApJ...675.1449R}), in addition to the IR lightcurve also suggesting that in spite of the HID classification,  \grs\ may not have strictly been in a plateau state. For this dataset we therefore use the average from MJD  53473.098 to 53473.25 of $44.9\pm3$ mJy, closest to the RXTE and CTIO observations.

\section{Constraints on input physical parameters}
\label{sec:grsparams}

In order to meaningfully compare broadband spectra, we will fit our
datasets using a model designed for simultaneous radio through X-ray
datasets. Historically the modelling of BHBs has focused on accretion
flow models of only the X-ray emission, of which Comptonising corona
models have been particularly successful.  However in recent years,
the evidence is mounting that the bipolar jet outflows found in the HS
contribute significantly across the broadband continuum.  For bright
transients, e.g. \citet{Russelletal2006,Russelletal2010} estimate up to 90\% of the NIR
and up to 100\% in the X-rays could be dominated by jet production. Other observations of
radio/IR/X-ray correlations
\citep{Corbeletal2000,Corbeletal2003,Coriatetal2009} can be
interpreted by either synchrotron or synchrotron-self Compton (SSC)
processes relating to the jets.  The outflow-dominated model of
\citet*{MarkoffNowakWilms2005} (hereafter: MNW05) is still the only model
that can fit the X-ray spectrum with the same precision as corona-only
models (ibid; Nowak et al., in prep.) while also fitting the radio
through IR bands from the same physical picture.  The MNW05 model has
already been applied successfully to many Galactic sources in the
``canonical" HS: the original paper features fits to Cyg X-1 and
GX~339-4, while a different data set of the latter, along with
observations of XTE~J1118+480 are fit in \citet{Maitraetal2009a}.
Further papers have explored applications to simultaneous broadband
data sets from GRO~J1655-40 and A0620-00
\citep{Migliarietal2007,Galloetal2007}, with other sources in
progress.  The results of all these applications has been the
discovery, perhaps not surprisingly, that the free parameters fit into
similar ranges for all stellar mass sources.  

Furthermore, the discovery of a Fundamental Plane of black hole
accretion \citep*{MerloniHeinzDiMatteo2003,FalckeKoerdingMarkoff2004}
supports mass-scaling accretion physics from stellar to supermassive
BHs, and thus would argue that the MNW05 should also apply to weakly
accreting AGN.  Confirming this, \cite{Markoffetal2008}
successfully fit several spectral energy distributions (SEDs) from the
supermassive BH M81$^*$ with parameters in the ranges found from BHBs.

There exists, therefore, a solid framework for the modelling of BHBs in
the canonical HS, against which to test fits of data from the
\grs\ plateau state. In this paper we do not consider whether MNW05 is correct but rather use it as a benchmark to test how far \grs\ fits the phenomenology of the hard state.

Whenever possible, we fixed values for model physical parameters in accordance with those found in other publications during the fitting process.  In the following subsections, we first discuss how we
obtained these values (presented in Table \ref{tab:frozen}), and then
briefly summarise the MNW05 model and our fitting methods before
presenting our results.

\subsection{Distance and Hydrogen Column Density}

The distance to \grs\ is still a matter of some debate. A first
estimate was derived in \citet{1994Natur.371...46M}. From a core
ejection they find a maximum distance of 13.7 kpc, under the
assumption that the ejection was intrinsically symmetric.
\citet{1995ApJS..101..173R} attempted to derive a more accurate
distance estimate by determining the kinematic distance from 21 cm
absorption spectra of atomic hydrogen along the line of sight to
\grs\ during a radio outburst, and find that it could be as far
away as 12.5$\pm1.5$ kpc. Later measurements of the $^{12}$CO($J=1-0)$
spectrum by \citet{1996A&A...310..825C} are consistent with this
distance. They also constrain the visual extinction to be
$A_v=26.5\pm1.7$ mag, corresponding to a total hydrogen column density
of $N_{\rm H}=4.7\pm0.2\times10^{22}$ cm$^{-2}$.  

If \grs\ resides at 12.5 kpc, it must be accreting above the
Eddington limit ($L_{\rm X-ray}\sim2.9\times10^{39}$ erg s$^{-1}$;
\citealt{2003astro.ph..6213M}, giving $L_{\rm X-ray}/L_{\rm
  Edd}\sim1.5$, using a mass of 14 M$_{\odot}$; see next Section).  A
slightly lower maximum distance of 11.2$\pm0.8$ kpc, derived from
proper motion studies of radio ejecta during four major events
\citep{Fenderetal1999}, would still give a ratio $L_{\rm
  X-ray}/L_{\rm Edd}\sim1.3$.  

However, this distance could be too large.
\citet{2004A&A...414..659C} measure $^{12}$CO($J=1-0)$ velocity
spectra of clouds along the line of sight, as well as of two nearby
HII regions, and re-evaluate the hydrogen column density to $N_{\rm
  H}=3.5\pm0.3\times10^{22}$ cm$^{-2}$.  This column gives a visual
extinction of $A_v=19.6\pm1.7$ mag, reducing the number intervening
molecular clouds, and arguing for a smaller distance to
\grs\ of 9.0$\pm$3.0 kpc (at 9 kpc the accretion rate estimate drops below Eddington), based on
evidence that \grs\ lies behind the molecular hydrogen cloud
G~45.45+0.06. This cloud is at $\sim7$ kpc
\citep{1998A&A...339..759F}, however \citet{2004A&A...414..659C} suggest that
this cloud belongs to a huge complex, located at a distance of 6
kpc, providing the above \emph{lower} limit to the distance (the upper limit of 12 kpc is from \citealt{Fenderetal1999}). However, the high column
density required \citep{1994Natur.371...46M,1995ApJS..101..173R}
suggests that another molecular cloud should be between us, therefore
\citet{Fenderetal1999} adopted the conservative distance of
11kpc.

Considering the assumptions made in \citet{2004A&A...414..659C} and
the resulting size of the errorbars, we have decided to follow
\citet{Fenderetal1999} in adopting a more conservative distance
of 11 kpc for this work. This distance is consistent with $N_{H}=3.5\pm0.3\times10^{22}$ cm$^{-2}$, which we believe is the correct value pertaining to the IR band, as well as the
larger hydrogen column density of 4.7$\times10^{22}$ cm$^{-2}$ from
\citet{1996A&A...310..825C}, that we use for our X-ray fits, %, as this value
%allows for enough hydrogen clouds to exist between us and the object
%to account for the observed column density.  The choice for a somewhat
%larger column density in the X-ray is further 
motivated by the works of
e.g. \cite{2002MNRAS.331..745K} (see Section \ref{sec:grsdisc}).
%, although we find their values less consistent with our data.

Because a distance of 11 kpc implies super-Eddington luminosities for
\grs, we also fit the data using the minimum distance of
6 kpc, in order to explore the importance of distance on the modelling
conclusions.  

\subsection{\grs\ Black Hole Mass}\label{sec:grsmass}

In order to constrain the mass of the compact object in a binary via
the mass function, the orbital period, inclination and the mass of the
companion (or the mass ratio) must all be known.  \citet*{2001Natur.414..522G} found an
orbital period $P_{\rm orb}$ of 33.5 days and a velocity amplitude
$K$=140$\pm15 {\rm km\,s^{-1}}$, giving a mass function
$f(M)=9.5\pm3.0 $ M$_{\odot}$. Assuming the binary plane is the same as that of the
accretion disc, the orbital inclination can be estimated from the
orientation of the jets. As no constant precession has been observed, the jets
can be assumed perpendicular to the accretion disc and orbital
plane. The exact inclination is however still open to debate, as it is
determined from the brightness and velocities of both the approaching
and receding blobs. At a distance of 11 kpc the inclination is
$i=66^{\circ}\pm2^{\circ}$ \citep{Fenderetal1999}. \citet{HarlaftisGreiner2004} were able to deduce a mass ratio $M_{\rm d}/M_{\rm X}=0.053\pm0.033$ from the rotational broadening of photospheric absorption lines of the donor star. Together these values give a mass for \grs\ of 14.0$\pm4.4$M$_{\odot}$ \citep*{2001Natur.414..522G}.  We use these as our fiducial values for the distance and mass.  For the 6kpc distance we calculate a smaller inclination of $i=50^{\circ}$ using \citet{Fenderetal1999}, and, using \citet{HarlaftisGreiner2004}
a mass of 23$M_{\odot}$.

\subsection{Properties of the Donor Star}
\label{sec:grsdonor}

A rough identification of the companion was performed by
\citet{2001ApSSS.276...31G}, by analysing absorption line features in
the NIR. They conclude that the companion is a class
III K-M giant and compare it successfully to a K2 III star. If the star
is indeed a K2 III giant, its temperature should be $T_d=4455\pm$190 K
\citep*{1999A&AS..140..261A}, and accretion proceeds via Roche-lobe
overflow \citep{2001ApSSS.276...31G}.  We use this temperature %as a reference value 
for the
simplistic blackbody spectral component representing the companion in
our spectral fits.

\begin{table}
\caption{Fixed physical parameters used in the fitting process, obtained from the literature. We omit the error bars, as they are not used in the fits. }
\label{tab:frozen}
\begin{tabular}{@{}lrrc}
\hline
parameter			&value		&units				& Reference\\
\hline
column density		&4.7			&10$^{22}$ cm$^{-2}$	& a \\
mass			&14 and 23	&M$_{\odot}$			& b \\
inclination			&66	and 50	&$^{\circ}$			& c \\
distance			&11 and 6		&kpc					& c, d \\
donor temperature	&4455		&K					& e \\ 
\hline
\end{tabular}
\\
References: a: \citet{1996A&A...310..825C}, b: \citet{2001Natur.414..522G}, c: \citet{Fenderetal1999}, d:  \citet{2004A&A...414..659C}, e: \citet*{1999A&AS..140..261A}\\
\end{table}

\section{modelling and Results}\label{sec:grsmodres}

\subsection{Description of model}

For all spectral fits we use the program \texttt{ISIS}
\citep{2000ASPC..216..591H} compiled with XSPEC version 12.3.1x
libraries \citep{1996ASPC..101...17A}. The model discussed below is
forward-folded through the X-ray detector response matrices, but
applied directly to the radio through NIR data. To account for the
additional uncertainties in the PCA response matrix, a 0.5\%
systematic error has been added in quadrature to all PCA data. As the
relative calibration of the PCA and HXTE instruments is not certain,
the normalisation factor for the PCA data is set to unity and tied to
the radio and NIR normalisations during the fits, while the HEXTE data
normalisation is left free to vary.

As the focus of our paper is on modelling the non-thermal spectrum, and
the exact stellar model needed for \grs\ is uncertain,
we use a simple blackbody to model the NIR, with the temperature fixed
as discussed in the previous Section.  This component serves to
account for the excess IR/optical flux level due to the star, so that
the overall model normalisation is correct.

In addition to the jet and the star we have an outer accretion flow in the form of a ''standard`` geometrically thin accretion disc \citep{shakurasunyaev1973}. This flow is parametrised by the radius of the inner and outer disc edges ($r_{\rm in}$,$r_{\rm out}$) and the temperature at the inner edge ($T_{\rm in}$). In the Schwarzschild geometry, the innermost stable circular orbit has
a radius of 6 r$_g$, but this radius can be reduced to 1 r$_g$ in the Kerr metric, when the BH is maximally spinning. The disc component serves to both fit spectral signatures of thermal emission from the accretion flow in the X-ray band, as well as contributing seed photons for inverse Compton scattering, but is only a weak element in comparison to the entire spectrum. 

We now summarise the physical parameters in our jet (see Table \ref{tab:grsresults}). One of the most important of these parameters is the jet luminosity ($N_j$). This factor scales with the accretion power at the inner edge of the accretion disc and represents the power going into the jet and acts as a normalisation factor. The power is equally divided to supply the internal and kinetic pressures. We assume that the kinetic energy is carried by cold protons, while the leptons do the radiating. The energy involved in the internal pressure goes into the particle and magnetic energy densities, with a ratio determined by $k$. $k=1$ equals equipartition and higher values indicate magnetic dominance. The initial velocity at the base of the jet, or ``nozzle" is the proper sound speed of an electron/proton plasma: $\beta\gamma c\sim0.4c$. The radius of the jet-base is also a free parameter, $r_0$ (expressed in in units of gravitational radius $r_g=GM/c^2$). The particles start of in a quasi-thermal (relativistic) Maxwellian distribution, the peak energy of which is determined by the electron temperature ($T_e$). Beyond the nozzle the jet is allowed to expand freely, or adiabatically, causing a longitudinal pressure gradient. This pressure gradient leads to a moderate acceleration of the jet along the direction of flow; the resulting velocity profile is calculated from the Euler equations and is roughly logarithmic, evening out at bulk speeds with Lorentz factors of $\Gamma\sim2-3$ (see e.g. \citealt*{Falckeetal2009}). In a segment of the jet located at some variable distance ($z_{\rm acc}$; also expressed in $r_g$) from the base, we assume a significant fraction (75 \%) of the leptons is accelerated into a power-law. The slope of the power-law particle distribution $p$ is also not pre-determined. In each jet segment after the acceleration front, the general shape of the distribution is assumed to remain the same, while the total lepton density decreases according to the adiabatic expansion. This is achieved assuming a continuous injection of ``fresh" energetic power-law distributed electrons ($N(E)dE\propto E^{-p}dE$) in each segment after the acceleration region. The jet is stopped at a distance $z_{\rm max}$ from the base. 

All fits are done using the following components: (1) The MNW05
steady-state outflow-dominated model that includes a multi-colour
blackbody accretion disc and a single blackbody for the companion
star; (2) an additive \texttt{Gaussian} line profile, with a line
energy left free to vary between 6 and 7 keV, and line width
$\sigma$ free to vary between 0 and 2 keV; Models (1)+(2) are either convolved with Compton
reflection from a neutral medium (\texttt{reflect};
\citealt{1995MNRAS.273..837M}) or multiplied with the smeared edge
model (\texttt{smedge}; \citealt{1991PhDT........55E}; using the ``standard" index for the photoelectric cross-section of $\sim-2.67$;), that accounts for relativistic smearing of the iron line, and multiplied with a photo-electric absorption model (\texttt{phabs}) to account for the interstellar medium. As the strength of an absorption feature (or edge)
is related to the strength of the according emission line, the line
width is in principle also related to the absorption edge width. Thus
when using the \texttt{smedge} model, we tie the width of the edge to
the width of the \texttt{Gaussian}. For the \texttt{reflect} model we
assume the viewing angle to correspond to the jet
inclination. Furthermore, although the \texttt{reflect} model already includes
an absorption edge, it is a sharp unsmeared edge, therefore we also tried fitting the data with both the
\texttt{smedge} and \texttt{reflect} models.

\subsection{Spectral fitting and results}\label{sec:grsfitting}

%After some initial experimenting, including either the \texttt{smedge}
%or \texttt{reflect} model, or both, we arrived at a set of recurring
%ranges in the models with satisfactory initial statistics for a
%broadband data set ($\chi_{\rm red}^2\sim2-3$). To quantify which
%model is best suited, we homogeneously fit the data from the same
%starting point, i.e. using the obtained initial ranges, we start the
%fitting routine with a high tolerance, and progressively reduce the
%tolerance, logarithmically halving it every step.  For both data sets
%we perform this procedure, with and without the NIR, for all three
%model combinations. The \texttt{smedge} model emerged as statistically
%favoured in all attempts. Adding a smeared edge to the \texttt{reflect}
%model improves the statistics of a reflection model somewhat, but as
%mentioned above, this is not truly self-consistent and the improvement
%appeared hardly significant. Therefore, we proceed to fit the data in
%greater detail with the \texttt{smedge} model.

\begin{table*}
\begin{minipage}{170mm}
\caption{Parameter ranges found in canonical black holes GX339-4 (\citealt{MarkoffNowakWilms2005}; Maitra, Markoff \& Brocksopp 2009), XTE J1118+480 (Maitra, Markoff \& Brocksopp 2009), Cyg X-1 \citep{MarkoffNowakWilms2005}, GRO J1655-40 \citep{Migliarietal2007} and A0620-00 \citep{Galloetal2007}, during the HS and best-fit parameters for the \grs\ plateau state. The error bars have been resolved at 90 percent confidence level. We failed to resolve the error bars for parameters listed in italics. $N_j$ is the jet normalisation, $r_0$ the nozzle radius, $T_e$ the temperature of the leptons as they enter the jet, $p$ the spectral index of the radiating particles, $k$ the ratio between magnetic and electron energy densities, $z_{acc}$ location of the particle acceleration region in the jet, $f_{\rm scat}$ 0.36/ ratio between scattering mean free path and gyroradius, L$_{\rm disc}$ and $T_{\rm disc}$ the luminosity and temperature of the accretion disc, $A_{\rm HXT}$ is the relative normalisation between HEXTE and PCA, $A_{\rm line}$, $E_{\rm line}$ and $\sigma$ are the normalisation, the energy and width of the iron line feature, and $E_{\rm edge}$ and $\tau_{\rm max}$ are the energy and the normalisation of the \texttt{smedge} model iron line edge (the iron edge width is the same as that of the \texttt{Gaussian}) and $\Omega/2\pi$ is the reflection fraction from the \texttt{reflect} model. }
\label{tab:grsresults}
\begin{tabular}{@{}lccccccccc}	
\cline{4-10}
			&			&range found	&\multicolumn{7}{c}{\grs}												\\
	%	\cline{4-7}\cline{9-11}	
			&			& in canonical	&\multicolumn{4}{c}{MJD 51367}									&&\multicolumn{2}{c}{MJD 53473}\\
	\cline{4-7}\cline{9-10}																																
variable		&	units		& BHBs		&{\sc smedge}			&{\sc reflect}			&{\sc smedge2}			&{\sc 6kpc}			&	&{\sc eltemp}			&{\sc synch}		\\
																																			
\hline		\hline																																	
$N_j$			&$10^{-1}$L$_{\rm Edd}$	&0.0034 - 0.71	&$4.78_{-0.00}^{+0.01}$		&$6.45_{-0.01}^{+0.00}$		&$9.91_{-0.00}^{+0.01}	$	&$2.22_{-0.01}^{+0.01}$		&	&$13.65_{-0.41}^{+0.00}$	&${\it 1.99}$		\\
$r_0$			&$GM/c^2$		&3.5 - 20.2	&$20.39_{-0.01}^{+0.01}$	&$9.30_{-0.00}^{+0.01}$		&$6.45_{-0.00}^{+0.00}$		&$4.94_{-0.01}^{+0.05}$		&	&$4.18_{-0.15}^{+0.24}$		&$3.6_{-0.6}^{+0.6}$	\\
$T_e$			&10$^{9}$ K		&20 - 52.3	&$9.21_{-0.01}^{+0.00}$		&$7.70_{-0.00}^{+0.01}$		&$8.45_{-0.00}^{+0.00}$		&$9.01_{-0.01}^{+0.06}$		&	&$3.94_{-0.02}^{+0.00}$		&${\it 8.38}$	\\
$p$			&			&2.1 - 2.9	&$2.30_{-0.06}^{+0.05}$		&$2.10_{-0.05}^{+0.03}$		&$1.84_{-0.00}^{+0.01}$		&$2.53_{-0.04}^{+0.06}$		&	&$2.14_{-0.03}^{+0.08}$		&$1.16_{-0.03}^{+0.03}$	\\
$k$			&			&1.1 - 7	&$692_{-1}^{+1}$		&$270_{-1}^{+1}$		&$28.9_{-0.1}^{+0.0}$		&$95_{-3}^{+2}$			&	&$707_{-265}^{+68}$		&{\it 200}		\\
$z_{acc}$		&$ 10^3 GM/c^2$		&0.007 - 0.4	&{\it 844}			&{\it 30}			&$20_{-1}^{+4}       $		&{\it 7.4}			&	&{\it 9}			&${\it 17}$		\\
$\epsilon_{\rm sc}$	&10$^{-4}$		&1.6 - 299	&${\it 0.48}$			&${\it 0.91}$	 		&$0.87_{-0.03}^{+0.02}$		&{\it 1.7}			&	&${\it 0.75}$			&$0.90_{-0.19}^{+0.22}$	\\
$r_{\rm in}$		& $GM/c^2$		&0.1-486	&$2.54_{-0.06}^{+0.00}$		&$2.91_{-0.02}^{+0.02}$		&$4.67_{-0.06}^{+0.04}$		&$0.62_{-0.01}^{+0.00}$		&	&$4.32_{-0.54}^{+0.53}$		&$3.6_{-0.5}^{+0.5}$	\\
$L_{\rm disc}$$^b$	&10$^{-1}$ L$_{\rm Edd}$&0.007 - 0.99	&$0.91$				&$1.26$	 			&$1.51$				&$0.73$				&	&$2.49$				&$1.82$			\\
$T_{\rm disc}$		&keV			&0.06 - 1.53	&$0.86^{-0.00}_{+0.01}$		&$0.83_{-0.00}^{+0.00}$		&$0.68_{-0.00}^{+0.01}$		&$0.88_{-0.05}^{+0.02}$		&	&$0.81_{-0.04}^{+0.03}$		&$0.82_{-0.06}^{0.04}$	\\
$A_{\rm HXT}$		&			&	-	&$0.91_{-0.01}^{+0.01}$		&$0.91_{-0.01}^{+0.00}$		&$0.92_{-0.01}^{+0.00}$		&$0.91_{-0.01}^{+0.01}$		&	&$0.90_{-0.01}^{+0.01}$		&$0.90_{-0.01}^{+0.02}$	\\
																																	
\hline																																	
$A_{\rm line}$		&10$^{-3}$		&	-	&$37_{-2}^{+5}$			&$80_{-2}^{+3}$			&$48_{-2}^{+2}$			&$29_{-2}^{+2}$			&	&$28_{-4}^{+8}$			&$25_{-3}^{+6}$		\\
$E_{\rm line}$		& keV			&-		&$6.31_{-0.05}^{+0.10}$		&$\dagger6.00_{-0.00}^{+0.06}$	&$\dagger6.00_{-0.00}^{+0.04}$	&$6.34_{-0.12}^{+0.09}$		&	&$6.35_{-0.09}^{+0.09}$		&$6.36_{-0.09}^{0.09}$	\\
$\sigma$		& keV			&-		&$0.81_{-0.06}^{+0.10}$		&$1.32_{-0.05}^{0.05}$		&$0.92_{-0.07}^{+0.06}$		&$0.67_{-0.14}^{+0.17}$		&	&${\it 0.22}          $		&${\it 0.07}$	\\
																																	
\hline																																	
$\Omega/2\pi$		&			&0.00 - 0.35	&-				&$\dagger0.30_{-0.01}^{+0.00}$	&-				&-				&	& -				& - 			\\
																																	
\hline																																	
$E_{\rm edge}$		& keV			&-		&$9.15_{-0.11}^{+0.15}$		&-				&$8.87_{-0.13}^{+0.14}$		&$9.08_{-0.16}^{+0.18}$		&	&$9.09_{-0.16}^{+0.25}$		&$9.13_{-0.25}^{+0.22}$	\\
$\tau_{\rm max}$	&$10^{-2}$		&	-	&$12_{-2}^{+1}$			&-				&$19_{-1}^{+1}$			&$12_{-2}^{+1}$			&	&$11_{-2}^{+2}$		 	&10$_{-2}^{+1}$		\\
\hline\hline																																
$\chi^2$/DoF		&			&		&113/147			&150/148			&166/147			&131/147			&	&59/81				&59/81			\\
			&			&		&	(=0.77)			&	(=1.02)			&(=1.13)			& (=0.89)			&	&(=0.73)			&(=0.73)		\\
\hline
\end{tabular}  
\end{minipage}
\begin{center}
{\bf MNW05 Canonical Parameter Ranges and \grs\ Best-Fit Parameters}\\
\end{center}
$\dagger$ {\it Value is pushing lower or upper boundary and is therefore not well constrained. }\\
$b$ Derived from model values for $r_{\rm in}$ and $T_{\rm disc}$.

\end{table*}

The results of all fits are presented in
Figures~\ref{fig:smedge}--\ref{fig:synch}, with respective parameters
listed in Table~\ref{tab:grsresults}. Below we discuss the individual
fits to both datasets in more detail.  

Initial fits revealed that, in the case of \grs , using \texttt{smedge} is statistically preferred to using \texttt{reflect}, to model the reflection features above $\sim10$ keV. Therefore, we proceed to fit the data in greater detail with the \texttt{smedge} model, but include a fit that employs \texttt{reflect} as this model has been used generally to model the disc reflection in previous works (see Section \ref{sec:grsparams}).

The spectral index of
the optically thick radio-NIR synchrotron emission is determined in
part by the internal jet plasma parameters such as the electron
temperature, but it is most sensitive to the Doppler beaming factor
(calculated from the inclination of the jets).  The closer the jet
axis aligns with the line of sight, the less inverse the observed radio
spectrum. Fixing the inclination according to observation and choosing a jet length for all fits of 10$^{16}$ cm, sufficient so that the slope through the radio data points is continuous, reveals a key difference between the two \grs\ plateau state datasets and canonical BHBs.  In order to avoid too
much excess in the NIR over the companion star BB, and to provide the
best NIR fit, the initial particle acceleration region in the jets
$z_{acc}$ needs to be at a distance of at least $\sim10^4$ $r_g$.
Therefore the post-acceleration region in the jets does not dominate
the X-ray emission below $\sim10$ keV, in contrast to canonical BHBs
in the hard state where acceleration is found to begin on the order of
10s of $r_g$.

In general we obtain values for electron energy index $p\sim2.2$ that
result in a significant synchrotron contribution to the X-rays below
$\sim 20-50$ keV, with a ratio of synchrotron to inverse Compton flux of
$\sim0.1$.  The %very soft 
steeply declining X-ray flux above $\sim 20$ keV %, compared to HS BHBs, 
is best fit by a dominant inverse Compton contribution from
the base of the jets.  With a much harder value of $p$ we could
conceivably fit more of the X-ray emission via synchrotron emission,
but only if the exponential decay shape plays a significant role (see
Figure \ref{fig:synch} for such a fit to dataset 2).  For very soft
values of the spectral index, synchrotron emission will not contribute
significantly, but fits without any synchrotron contribution to the X-ray at all are not statistically favoured. 

The exact spectral shape and normalisation in the post-acceleration
synchrotron component (fitting the radio data) are very dependent on
the value of $p$.  When calculating a local statistical minimum, the
fitting routine naturally favours the plethora of X-ray data points over
the few radio data points, in finding the confidence limits for $p$,
explaining non-negligible systemic residuals in the radio.  

For clarity we note that below we will use {\sc small caps} to refer to fits and will continue to use the \texttt{typewriter} font to refer to models.

\subsubsection{Dataset 1}

\begin{figure}
\includegraphics[width=84mm]{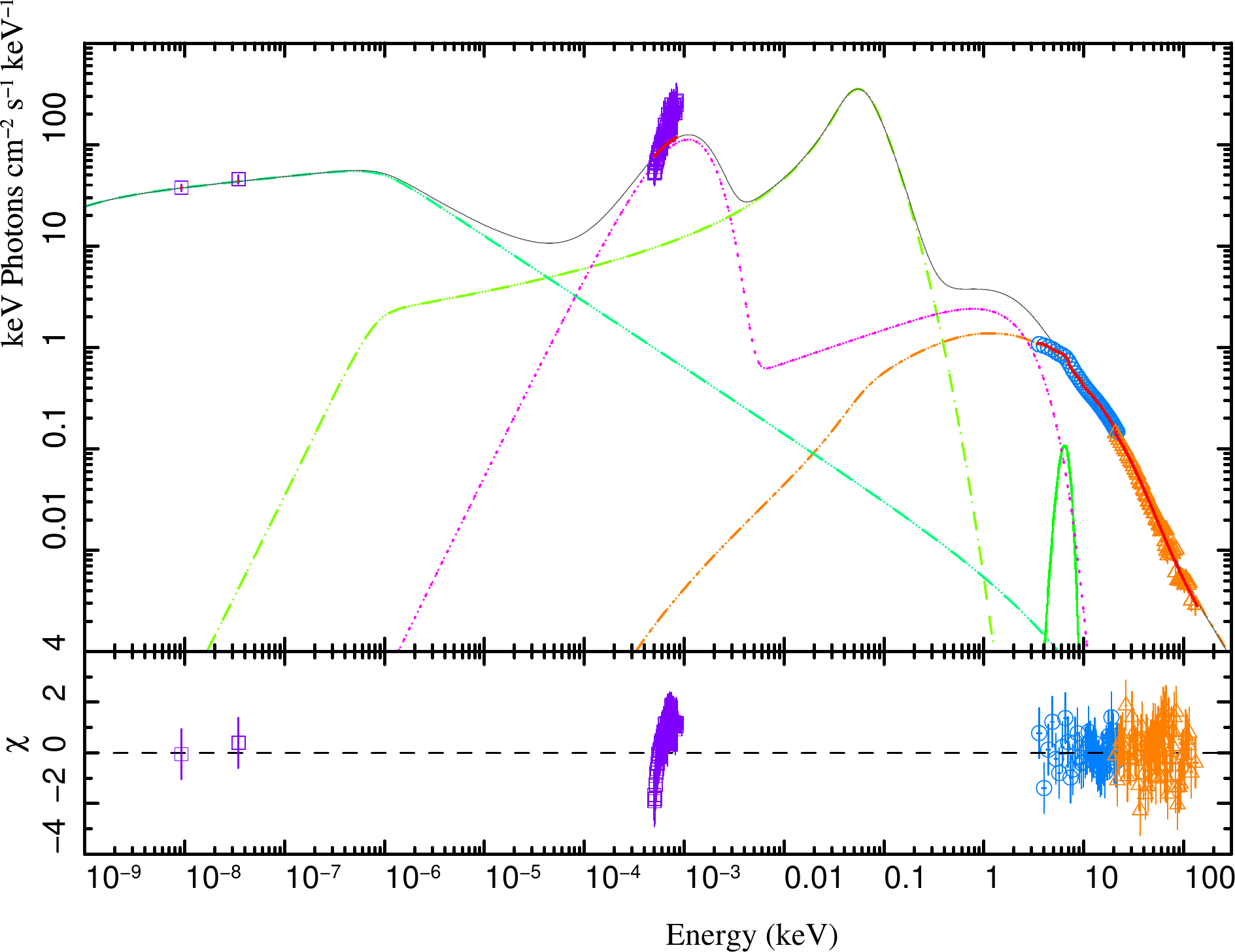}
\includegraphics[width=84mm]{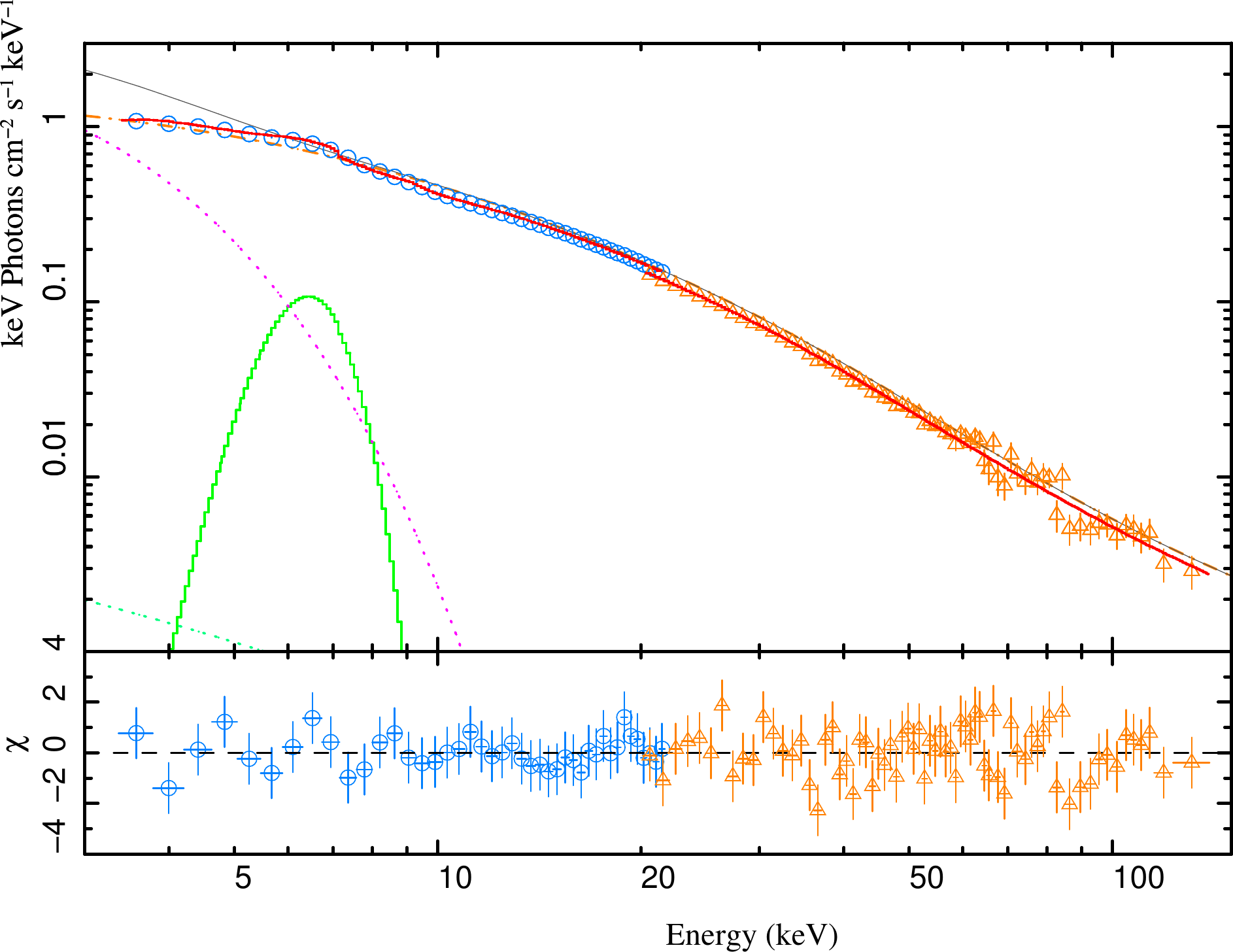}
 \caption{Multi-wavelength (top) and X-ray band only (bottom) best fit ($\chi_{\rm red}^2\sim$1.16) model spectrum (cf. Table \ref{tab:grsresults}, {\sc smedge} column), using MNW05+\texttt{Gaussian}+\texttt{smedge}. Individual contributions from the MNW05 model are also shown: The light-green dashed curve is the pre-shock synchrotron contribution. The dark-green dash-dotted curve is the post-shock synchrotron. The purple dotted curve represents the thermal contributions from the stellar blackbody (below $\sim10^{-2}$ keV) and the accretion disc. The orange dashed-dotted line above $\sim10^{-3}$ keV represent the Compton-upscattered stellar blackbody and accretion disc seed photons and the Synchrotron Self-Comptonisation (SSC) of the pre-shock synchrotron. The solid grey line is the total MNW05 model spectrum originating from the jet, the accretion disc and the companion, however not forward-folded through the detector response matrices and without iron line or reflection contributions or absorption due to the interstellar hydrogen column density or the \texttt{smedge} model. The red solid lines through the data points shows the model flux including all these features. The iron line complex, modelled by a \texttt{Gaussian} is shown by the thick green curve near 6.4 keV.}
 \label{fig:smedge}
\end{figure}

\begin{figure}
\includegraphics[width=84mm]{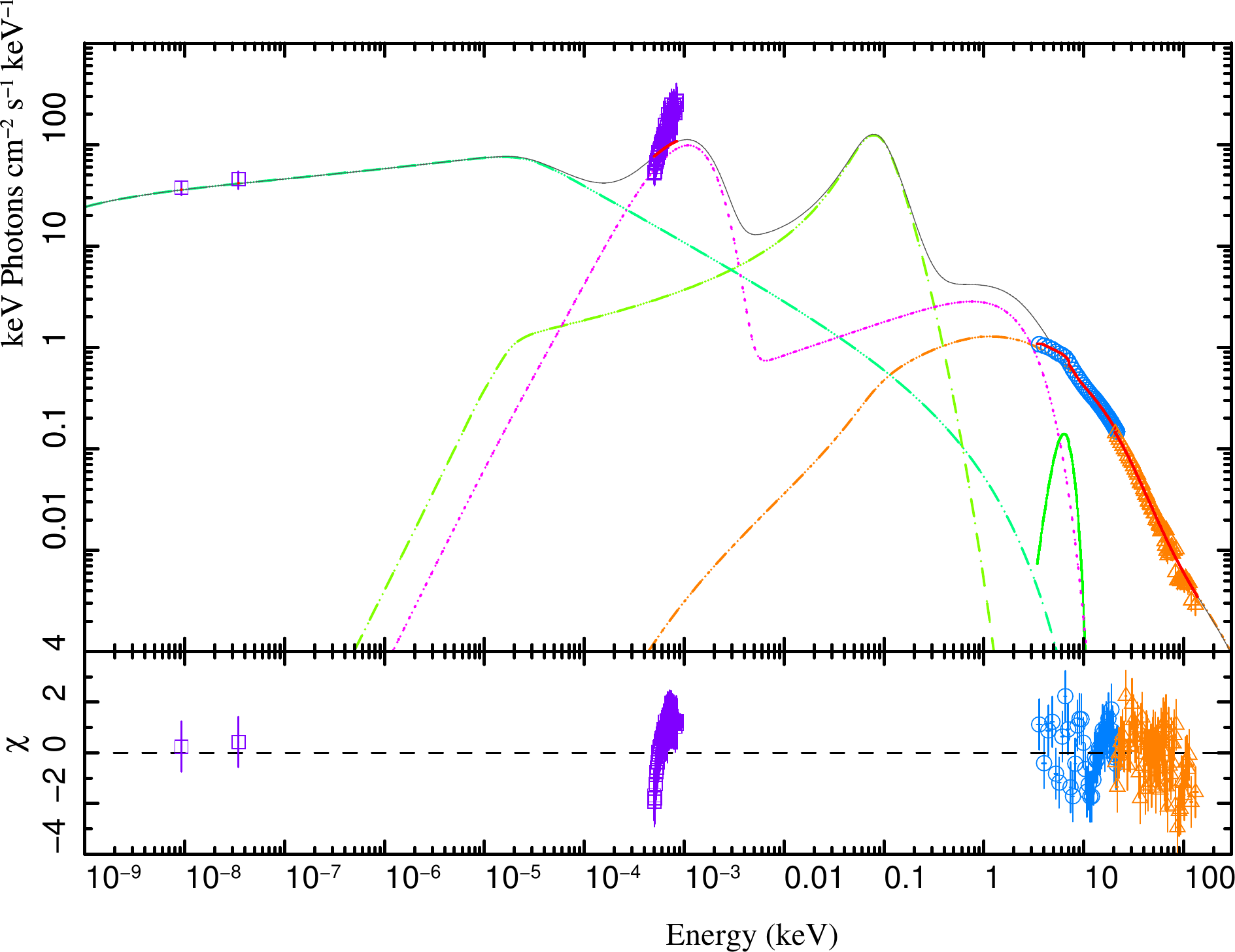}
\includegraphics[width=84mm]{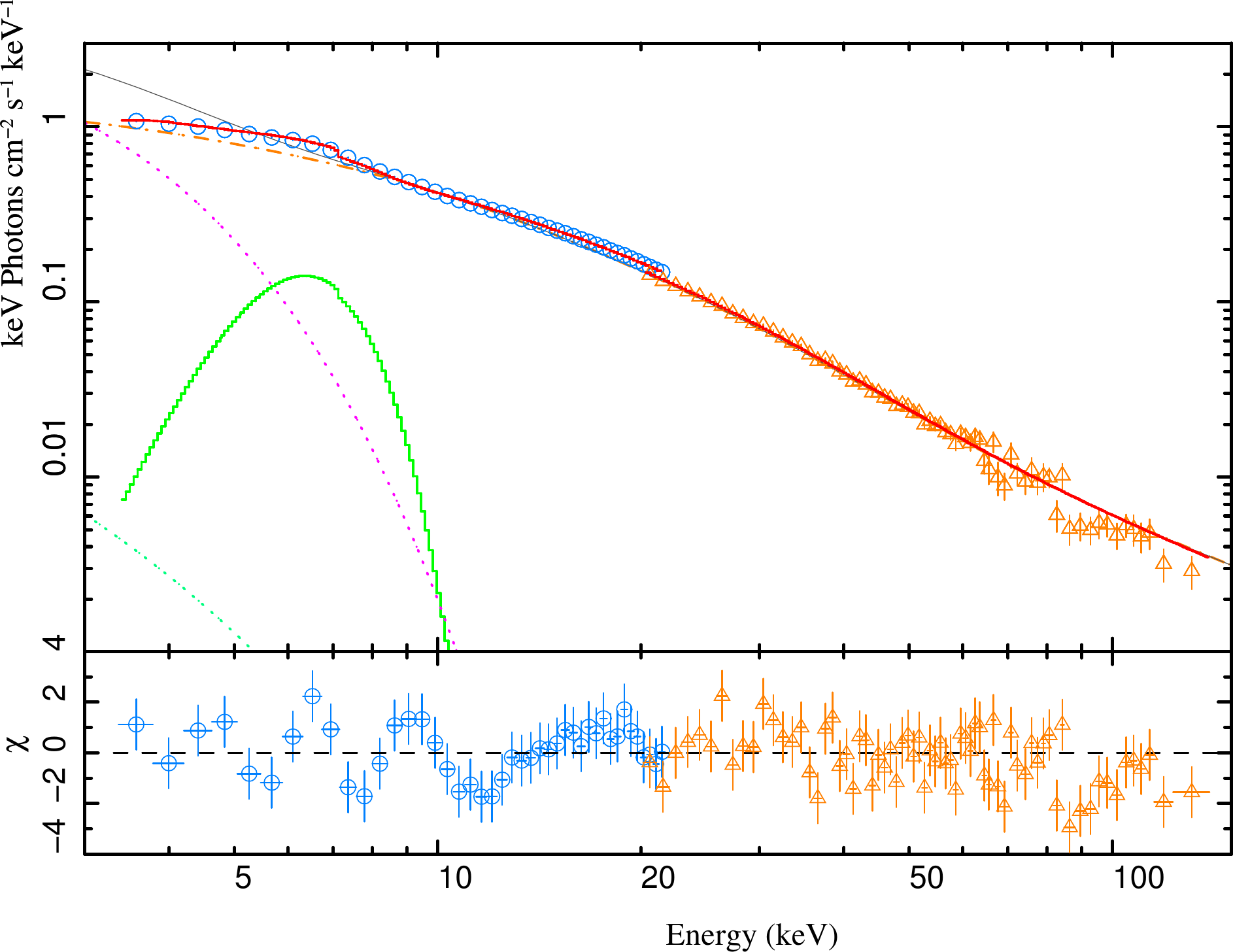}
 \caption{Multi-wavelength (top) and X-ray band only (bottom) best fit ($\chi_{\rm red}^2\sim$1.48) model spectrum (cf. Table \ref{tab:grsresults}, {\sc reflect} column), using MNW05+\texttt{Gaussian}+\texttt{reflect}.}
 \label{fig:refl}
\end{figure}

\begin{figure}
\includegraphics[width=84mm]{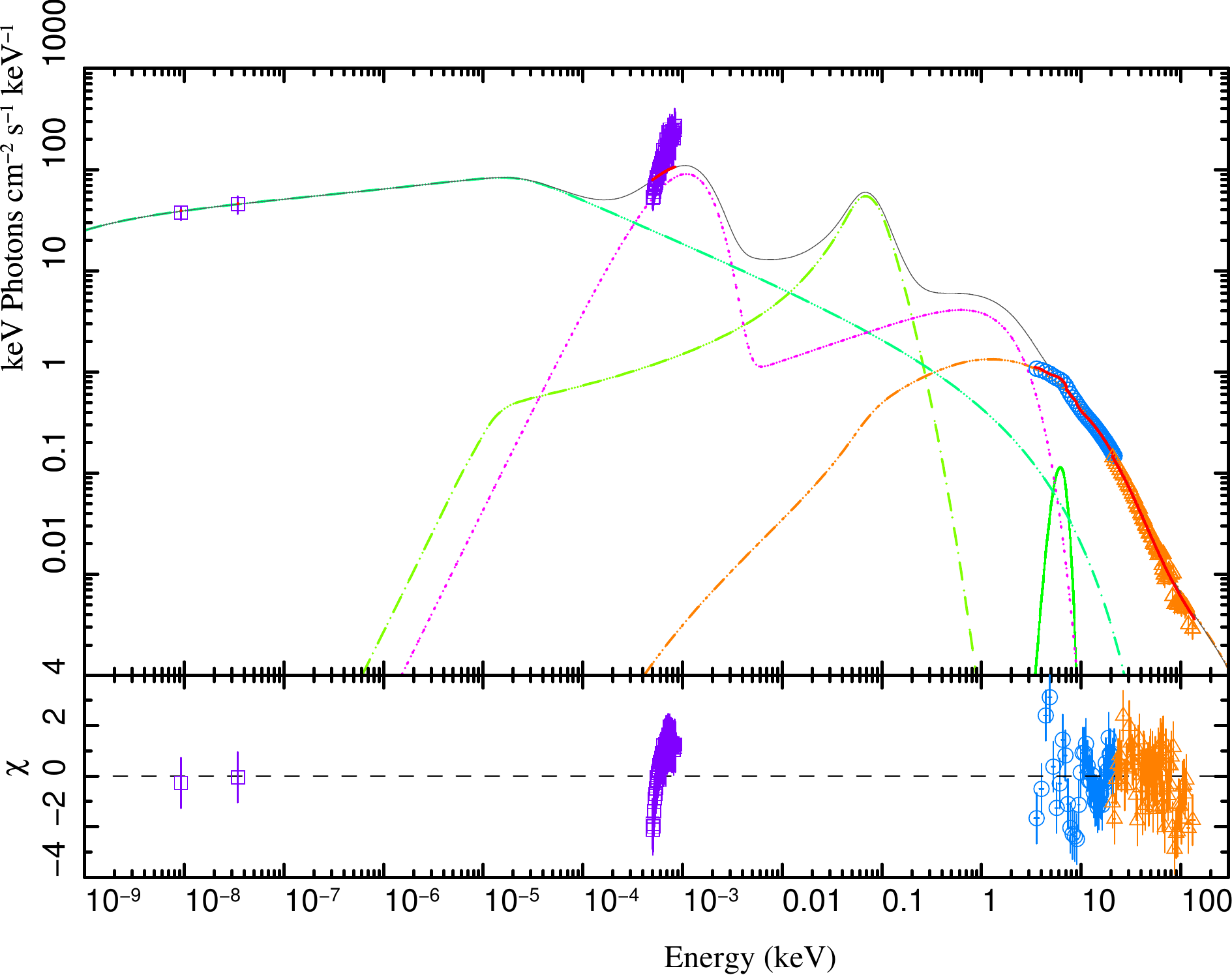}
\includegraphics[width=84mm]{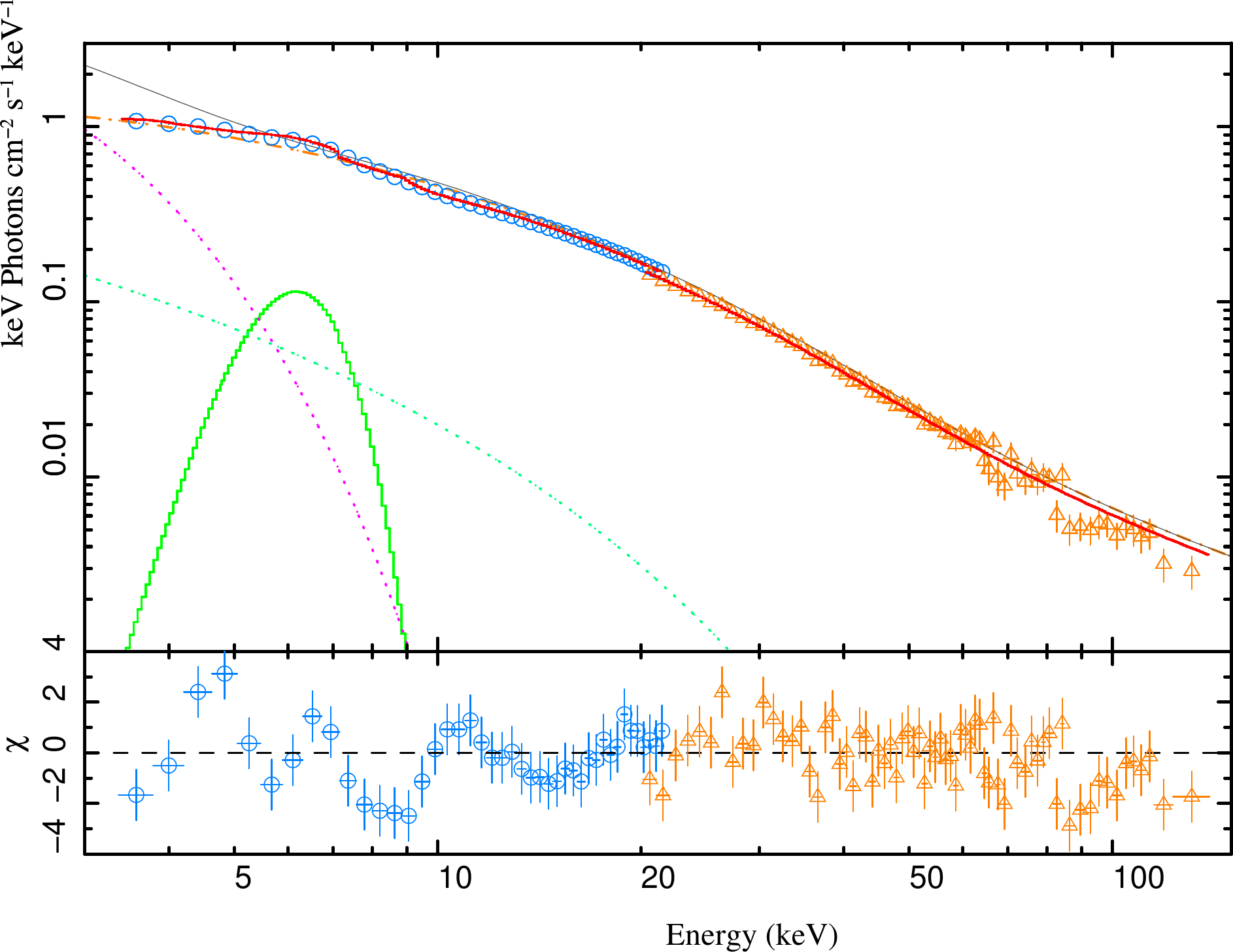}
 \caption{Multi-wavelength (top) and X-ray band only (bottom) best fit ($\chi_{\rm red}^2\sim$1.44) model spectrum (cf. Table \ref{tab:grsresults}, {\sc smedge2} column), also employing MNW05+\texttt{Gaussian}+\texttt{smedge}, but with a reduced magnetic dominance ($k\sim30$ in stead of $\sim550$, see table \ref{tab:grsresults}) .}
 \label{fig:lowequip}
\end{figure}

\begin{figure}
\includegraphics[width=84mm]{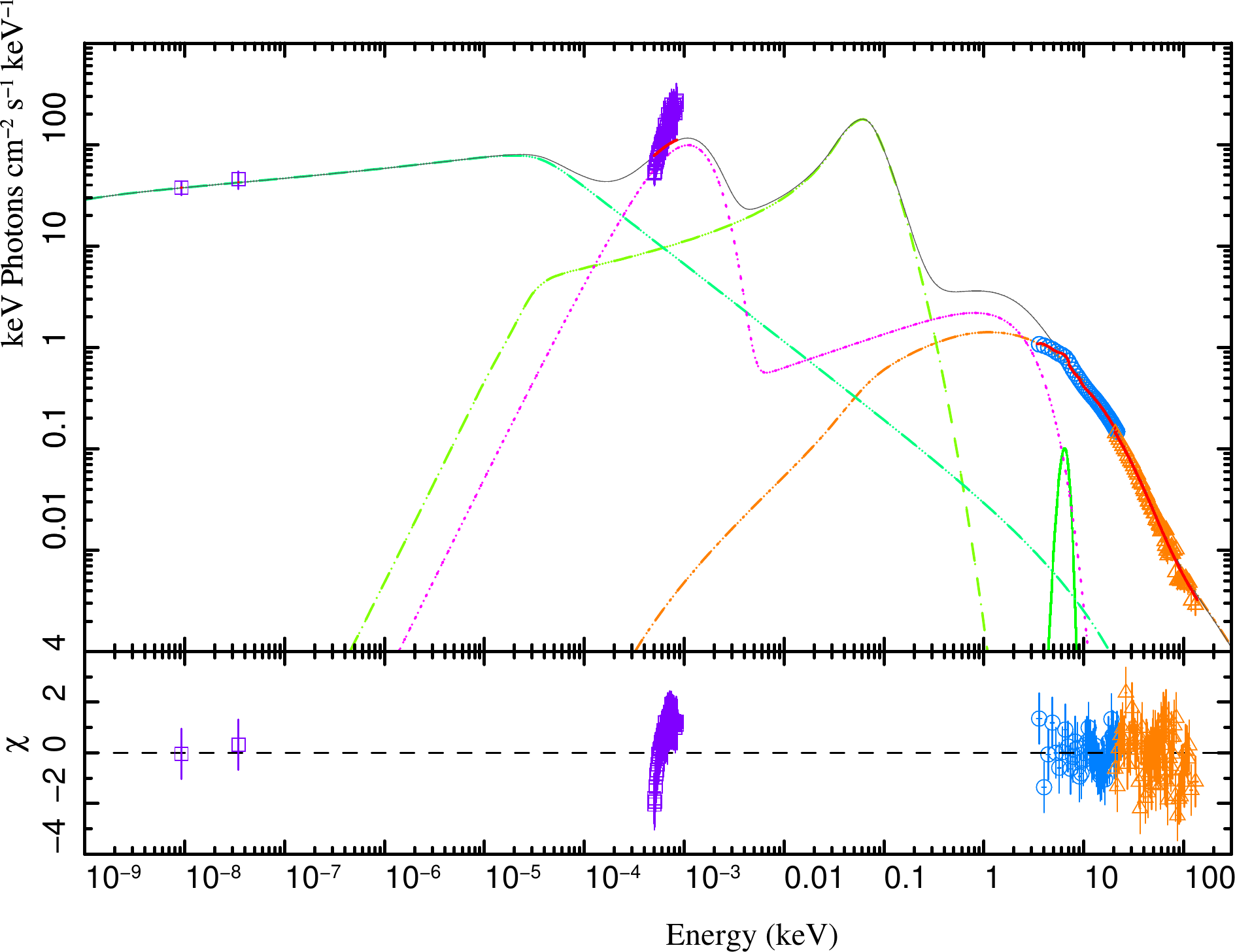}
\includegraphics[width=84mm]{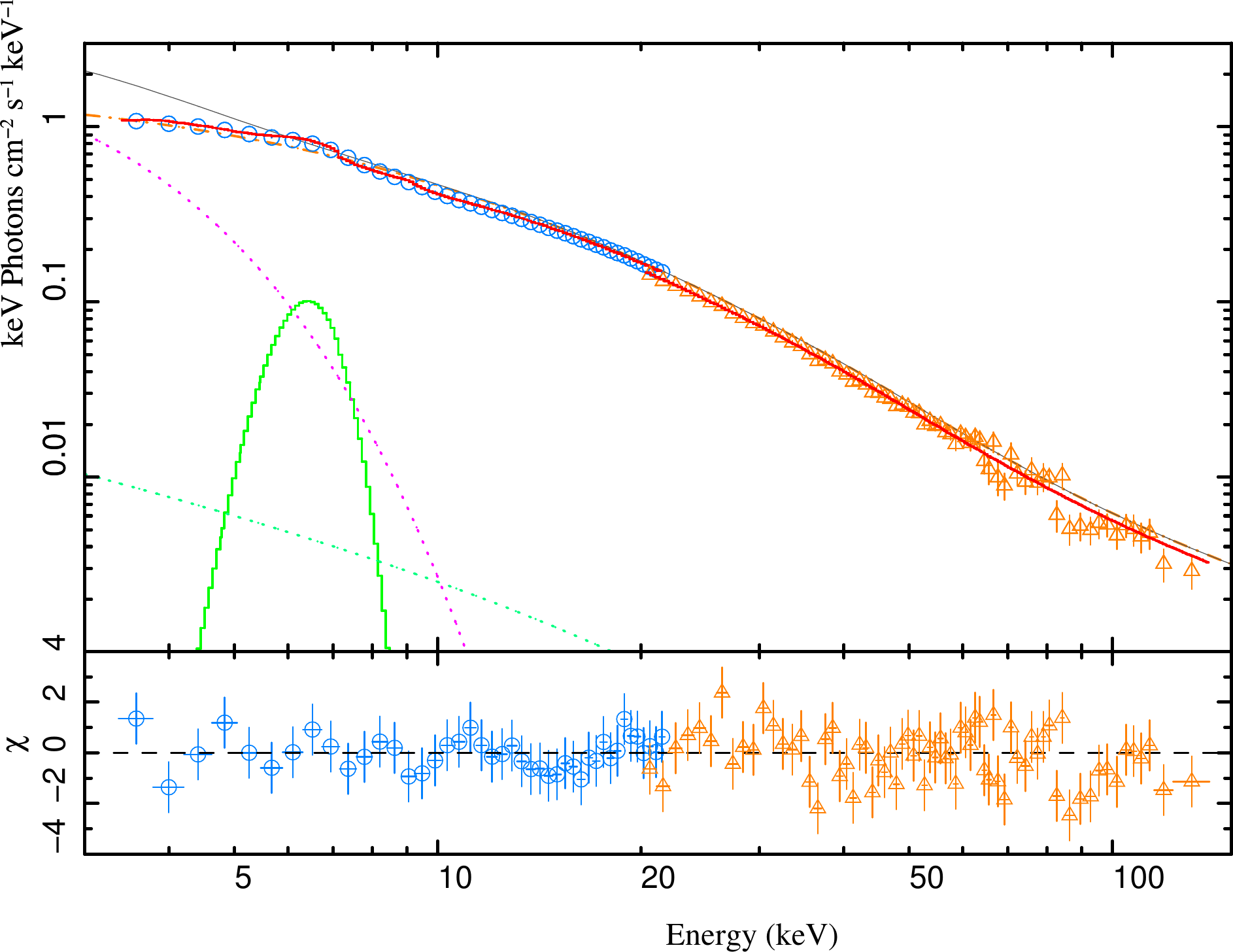}
 \caption{Multi-wavelength (top) and X-ray band only (bottom) best fit result ($\chi^2_{\rm red}\sim1.01$) spectrum (cf. Table \ref{tab:grsresults}, {\sc 6kpc} column), for the 1999 data set, using MNW05+\texttt{Gaussian}+\texttt{smedge} and employing a fixed distance of 6 kpc.}
 \label{fig:6kpc}
\end{figure}

The best fit models for dataset 1 are shown in Figures
\ref{fig:smedge} - \ref{fig:6kpc} (corresponding to Table
\ref{tab:grsresults}, {\sc smedge}, {\sc reflect}, {\sc smedge2} and {\sc
  6kpc} column respectively; see below for explanation of these fit names). The fits are in general statistically
good, with a large contribution to the residuals coming from the
inadequacy of a simple black body to fit the NIR data (see Figure
\ref{fig:ir}). Completely removing the NIR data from the best fit
{\sc smedge} models an improvement in the $\Delta\chi^2_{\rm red}$ of
$\sim0.08$, to $\chi^2_{\rm red}=0.69$. 

Our best fit {\sc smedge}, employing the \texttt{smedge} model also shows the most extreme behaviour in terms of the distribution of the energy budget (evident from the value for the magnetic dominance $k$) and the acceleration front distance $z_{\rm acc}$. In addition this fit is likely to suffer from pair production (see Table \ref{tab:ppp} and Section \ref{sec:grsdisc}). Hence we explore several fits with less extreme physical properties, including two more fits employing a \texttt{smedge} model, ({\sc smedge2} and {\sc 6kpc}) and one fit using the
\texttt{reflect} model ({\sc reflect}). The {\sc smedge} and {\sc smedge2}
fits are different local minima that primarily differ from each other in the distribution of the
energy budget, most importantly the latter fit has a reduced value for the magnetic dominance $k$ and particle distribution index $p$. For {\sc 6kpc} we reduced the distance to \grs\ to 6 kpc (see below).
These fits clearly show over an order of magnitude decrease in $z_{\rm acc}$ and a factor $2-25$ reduction in $k$.
Moreover, pair production will be less important. We note here that fixing $z_{\rm acc}$ to $2\times10^4$ $r_g$ in the {\sc smedge} fit results in only a minor increase in $\chi^{2}_{\rm red}$ of $0.05$, while producing virtually the same model parameters. This suggests that in this case our model does not converge and shows artefacts from the fitting routine that tries to find the absolute local minimum (within the tolerance limit), although there is no real physical basis for this insignificant improvement. Hence for the {\sc smedge} fit we will ignore the value for $z_{\rm acc}$ but include the other parameter values in our analysis.

The steeper optically thin synchrotron spectrum of {\sc smedge2} could be reconciled with the steeper $p$ found
in canonical BHBs by assuming that both share such a hard injected power-law, and 
considering the effect of cooling on the spectrum. We
do not consider cooling explicitly here, but are investigating such issues in a
separate work.  Because of the diminishing magnetic field strength
along the jets, the cooling (dominated by synchrotron losses) is
negligible. %if a maximum energy particle does not reside in a jet segment long enough to radiate most of its energy under the influence of the magnetic field.   
The break energy where the particle distribution steepens in
``particle space" shifts according to $E\propto B^{-2}\propto r^2$
\citep{Kardashev1962}. Thus the further along the jet we go, the less 
important cooling becomes, and thus the largest contribution to
the higher frequencies from synchrotron radiation is from the
first acceleration zone.  Because this zone occurs almost
two orders of magnitude further out in \grs\ compared to BHBs, this
effect could explain why we are seeing the uncooled injected spectrum
rather than the cooled spectrum, expected to be steeper by 0.5.  

From the arguments outlined in the previous Section it is clear that
the true distance to \grs\ could be much smaller than the 11
kpc adopted for most of our fits.  In an attempt to explore the effect
of distance, we performed one fit on dataset 1 with the distance to
\grs\ reduced to 6 kpc (see Figure \ref{fig:6kpc} and Table
\ref{tab:grsresults}, {\sc 6kpc} column). The modifications result in a
similar $\chi_{\rm red}^2$ but some values are closer to what
we have come to expect for the HS in other black holes. 
%Although the values for $f_{\sc}$ are not constrained, 
Firstly we find $z_{acc}$ to be the smallest of all fits. This can be understood from the smaller inclination of $50^{\circ}$ corresponding to the smaller distance, resulting in a less inverse radio spectrum and hence a lower flux level at the IR break. In addition the jet luminosity $N_j$ and the partition parameter $k$ are closer to
values found for other BHBs.  With $N_{\rm j}\sim0.22 L_{\rm Edd}$ and $k\sim90$
they are only a fraction of the values found for most 11
kpc fits. Also the smaller nozzle radius $r_0\sim5$ is more typical of
the usual HS value. One possible interpretation of these results is
that \grs\ might in fact be closer than the conservative value
usually taken in the more recent literature.

\subsubsection{Dataset 2}

\begin{figure}
\includegraphics[width=84mm]{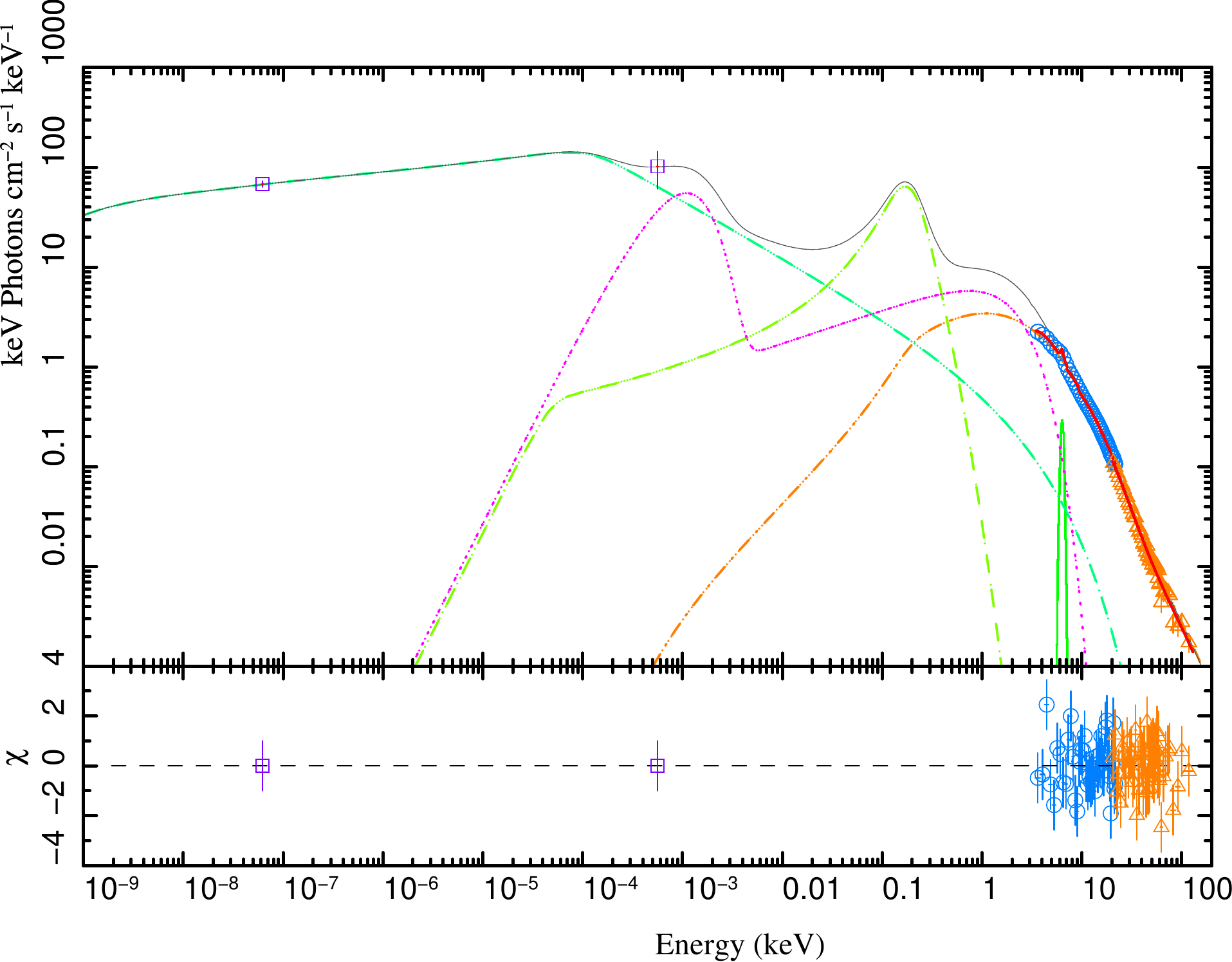}
\includegraphics[width=84mm]{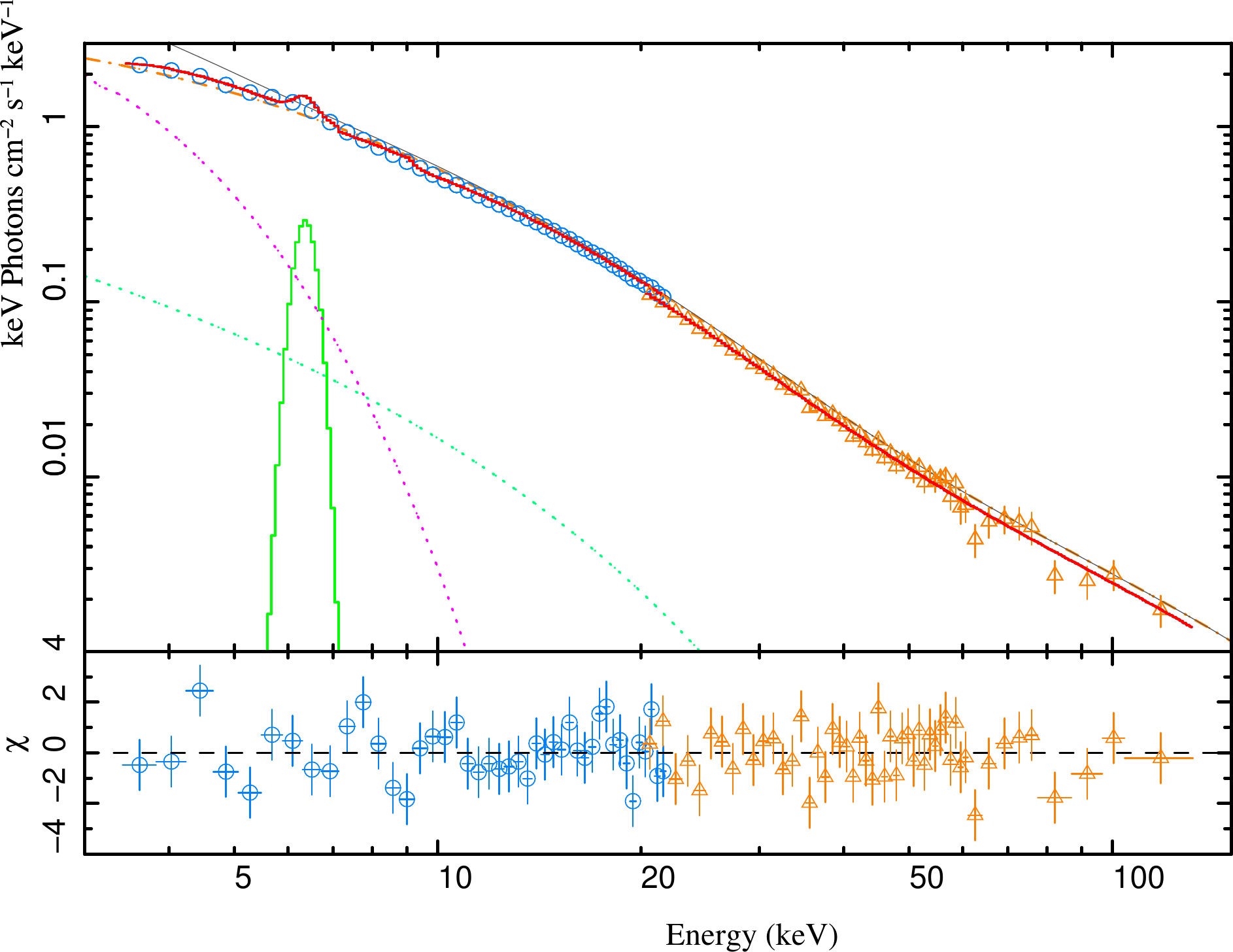}
 \caption{Multi-wavelength (top) and X-ray band only (bottom) best fit result ($\chi^2_{\rm red}\sim1.0$) spectrum, for the 2005 data set, using a reduced low electron temperature $T_e\sim4\times10^{9}$ K (cf. Table \ref{tab:grsresults}, {\sc eltemp} column). }
 \label{fig:eltemp}
\end{figure}

\begin{figure}
\includegraphics[width=84mm]{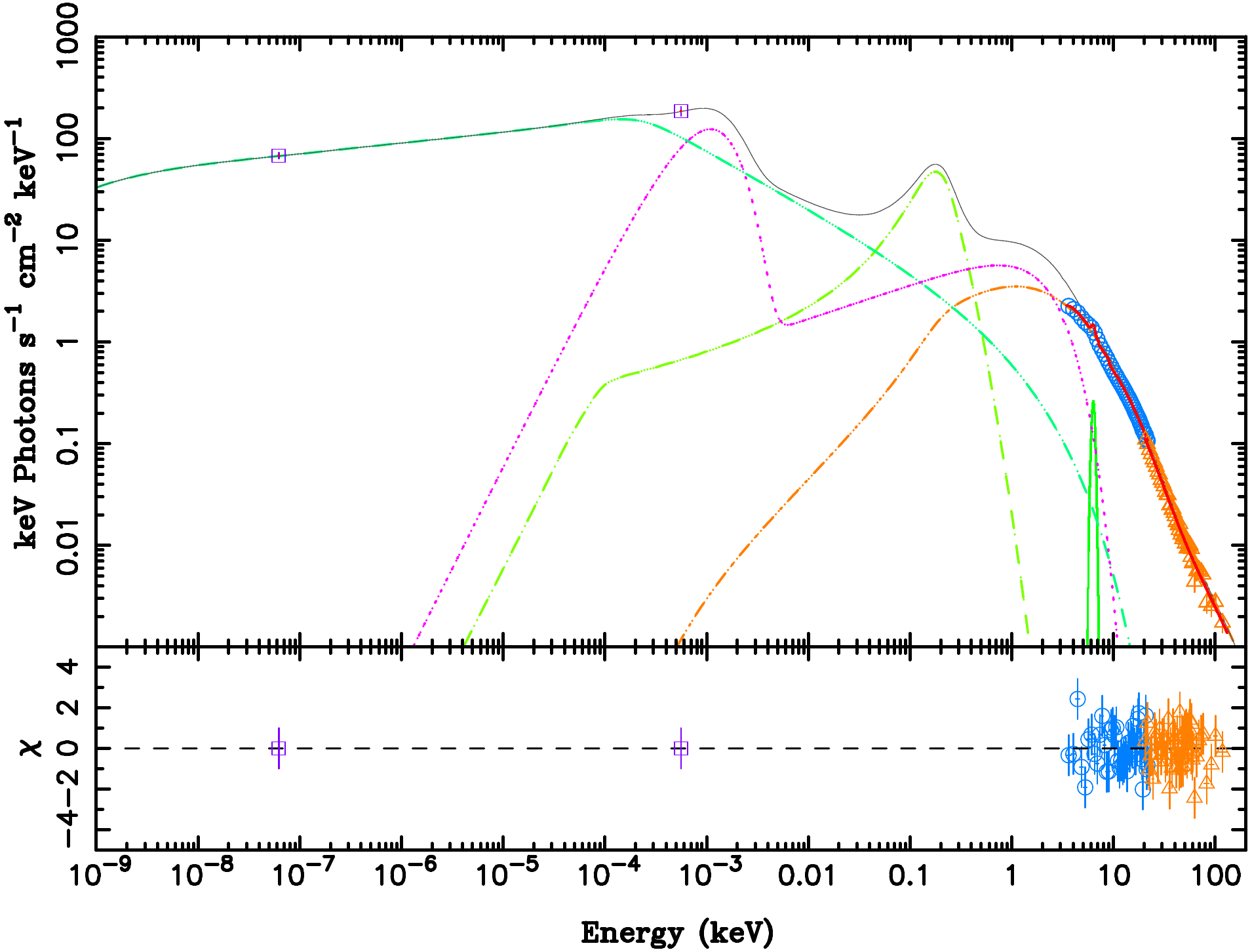}
\includegraphics[width=84mm]{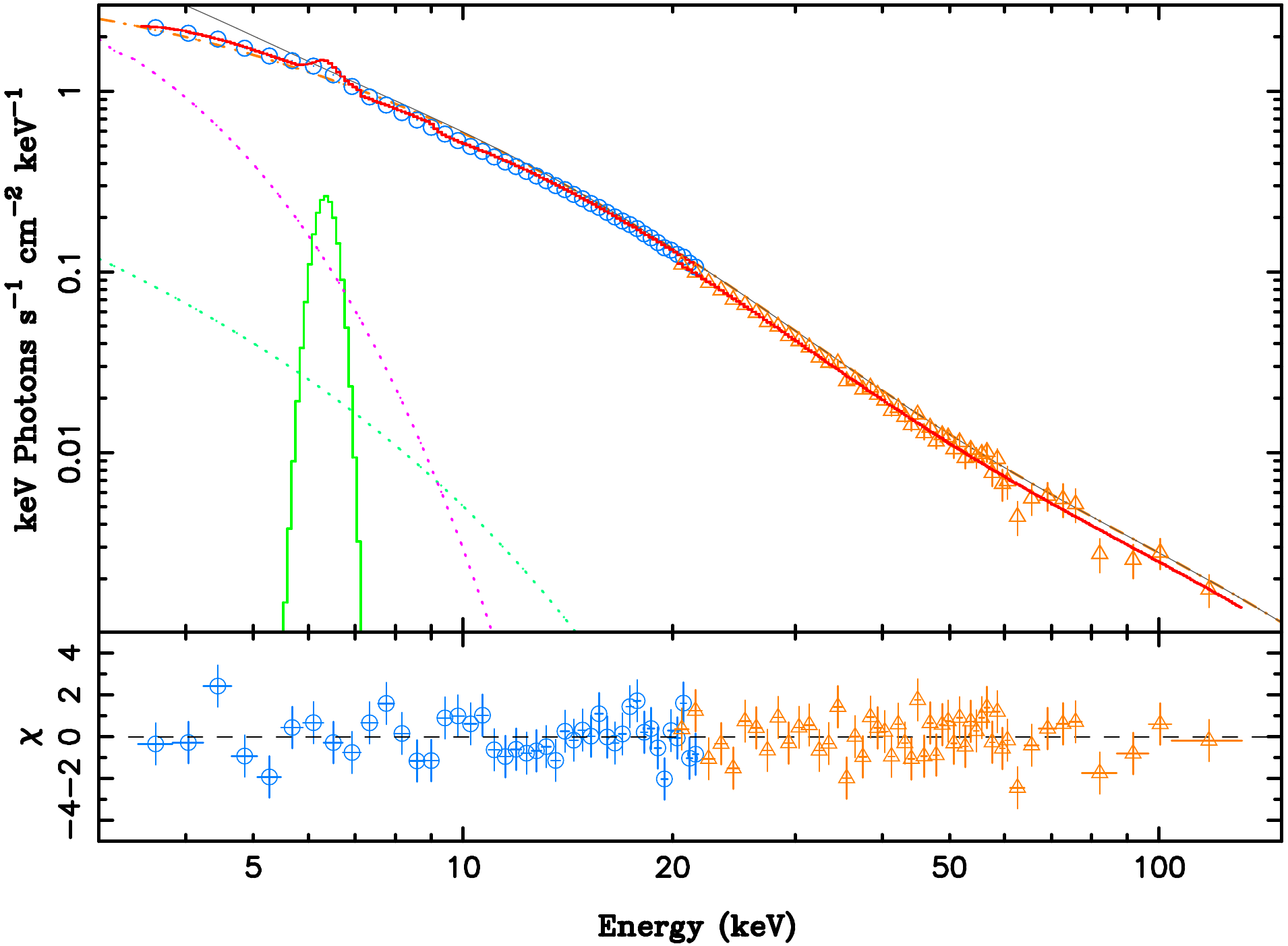}
 \caption{Multi-wavelength (top) and X-ray band only (bottom) best fit result ($\chi^2_{\rm red}\sim1.15$) spectrum, for the 2005 data set, using the synchrotron cut-off to fit the X-ray (cf. Table \ref{tab:grsresults}, {\sc synch} column).}
 \label{fig:synch}
\end{figure}

The best fit models for dataset 2 are shown in Figures
\ref{fig:eltemp} and \ref{fig:synch} (corresponding to Table
\ref{tab:grsresults}, column, {\sc eltemp} and {\sc synch}
respectively).  {\sc eltemp} offers the best comparison -- in terms of what component fits what feature -- to dataset 1, and owes its name to its low electron temperature $T_{\rm e}$, while {\sc synch} is dominated by the post acceleration synchrotron component.

Fitting dataset 2 with similar parameter values (relating to the
energy budget and assumed geometry) as dataset 1 yields very poor
statistics (best $\chi^2_{\rm red}\sim30$). 
%Even using both the
%\texttt{smedge} and \texttt{reflect} models, despite the
%aforementioned inconsistency, does not improve matters
%much. Furthermore the constant that determines the HEXTE normalisation
%($A_{\rm HXT}$) is pushing on our imposed lower boundary ($0.8$) in an attempt to fit the hard X-ray spectra via a renormalization rather than a slope change (see Figure 10), 
%The fact that this happens
%indicates that the same approach as to dataset 1 is not able to
%provide a sufficient combination of components to fit the must softer
%dataset 2. 
Trying to model the steep X-ray spectrum, mainly employing
the exponential decay of the multicolour disc fails, because a higher
disc contribution offers more (soft) seed photons. These photons are
up-scattered and create a Compton tail in the hardest part of the
spectrum where they overestimate the observed flux.  
%The {\sc compare} fit is
%shown for comparison with dataset 1, using similar values for the
%parameters, however it has extremely poor statistics and therefore no
%conclusions are drawn on the basis of this fit.

We therefore searched the parameter space for any other solution that reduces the flux of
Comptonised high-energy photons by reducing the electron temperature
$T_e$. A lower temperature electron distribution would on average not
up-scatter the disc seed photons to equally high energy. We find that
if we allow $T_e$ to evolve freely, we end up with a very good fit
($\chi^2_{\rm red}\sim0.7$), however the final temperature is rather low. At $\sim
4\times 10^9$ K the bulk of the electrons would be sub-relativistic. The use of a
relativistic Maxwell-Juttner distribution, as done in the MNW05
model, would no longer be fully justified, as the majority of the particles
in the thermal distribution would have $\gamma=1$. However this approach effects a very peaked and steep pre-shock
synchrotron component. This distribution of seed jet-base photons
allows for a Comptonised photon distribution that is steep enough to
fit the X-ray data up to 50 keV. Above 50 keV the Comptonised disc
photons harden the model spectrum just enough to fit the entire range
of HEXTE data.

An alternate possibility for fitting dataset 2 - with an increased
electron temperature - is found when we increase the ratio of
synchrotron to inverse Compton emission.  With
synchrotron dominating below $\sim 30$ keV
(see Figure \ref{fig:synch} and Table \ref{tab:grsresults}, {\sc synch}
column) and cutoff at about 1 keV, the subsequent
exponential decay can approximate the X-ray features rather well, when
combined with an accretion disc peaking at the same energy (Figure
\ref{fig:synch}).  In general we feel that, given the unknowns in the
exact shape of the particle distribution around the cutoff, it is
undesirable to rely on this in general during the fits.  However it is
worth noting that it provides a very good description of the data
($\chi^2_{\rm red}\sim0.7$). Fitting with the synchrotron cutoff
requires a very hard synchrotron spectrum ($\Gamma\sim1.2$). 
% but delivers a credible HEXTE normalisation constant $A_{\rm HXT}$.
Normally such a hard spectral index would be expected only in
ultra-relativistic sources \citep[e.g.][]{1988MNRAS.235..997H}.  As for the {\sc smedge2} fit, the relatively shallow spectrum could, however, again be (partly) reconciled with the canonical values for $p$, considering the effect of cooling on the spectrum.

\subsection{Stellar companion spectrum}

While the predominant temperature of a K2 star is 4455K, the dataset 1 fits show a systemic discrepancy that could be resolved increasing the blackbody temperature (and normalisation). While the temperature of 4455 K is the predominant
value for a solitary class 2 star of this type, clearly this is not the case for the companion of \grs.  

Although the physical properties of stars under these conditions are
still largely undetermined, \citet{2001ASSL..264..125K} resolved
discrepancies in the effective temperatures of donor stars in high
mass X-ray binaries, of 10-25\% higher than expected from their
spectral classification.  Increasing the temperature of our blackbody
by 25\% to 5500 K results in an insignificant improvement in fit by $\Delta\chi_{\rm
  red}^2=0.03$.  
  
  Letting the temperature and normalisation
of the blackbody in our best-fit result {\sc smedge} free to vary improves the $\chi_{\rm red}^2$ to $81/165=0.5$, but yields an extremely high blackbody temperature of 1.33$_{-0.04}^{+0.14}\times10^{5}$ K. Such an
exceedingly high temperature indicates that other IR contributions
than photospheric emission from the secondary are
likely. \citet{1996A&A...310..825C} already concluded the same from
the rapid variations and spectral shape they observed in this band. In
particular irradiation of the companion star and/or the disc for a source as bright as \grs\
should be quite significant, but the former is not accounted for in detail in our modelling, while the latter is not accounted for at all. Using an irradiated disc+jet model, \citet{Maitraetal2009a} obtained disc temperatures for XTE~J1118+480 in the order of a few tens of thousands Kelvin, in agreement with e.g. \citet{Hynesetal2006}. Our obtained temperature of $\sim1.3\times10^{5}$ K seems rather high, however XTE~J1118+480 accretes at rates $\lesssim10\%$ $L_{\rm Edd}$, while for \grs\ this is $\sim L_{\rm Edd}$. 
Other possibilities for IR contributions include optically thick free-free
emission (although usually not observed at such high energies), or
optically thin free-free radiation from an X-ray driven accretion disc
wind \citep*{2003A&A...404.1011F,CastroTiradoetal1996}.

%\begin{figure}
%\includegraphics[width=84mm]{fig10a.pdf}
%\includegraphics[width=84mm]{fig10b.pdf}
 %\caption{Multi-wavelength (top) and X-ray band only (bottom) best fit result ($\chi^2_{\rm red}\sim39$) spectrum, for the 2005 data set, using similar values as employed for dataset 1. Note the decreased resolution on the residual axis. Furthermore the residual of lowest energy X-ray point is left out so the others can be better seen in the plot.}
 %\label{fig:bestfit}
%\end{figure}

\begin{table}
\caption{Relative importance of pair production processes at the base of the jet, for the statistically good fits in table \ref{tab:grsresults}. $n$ is the lepton number density per cm$^{-3}$, $\dot{n}_{\rm pa}$ and $\dot{n}_{\rm pp}$ are the pair annihilation rate and production rate per cm$^{-3}$s$^{-1}$, respectively. In order to compare the production rates to the lepton density, they have been multiplied by an estimate for the \emph{residence time}, given by $r0/c$. If $n\gg$ Max($\dot{n}_{\rm pa},\dot{n}_{\rm pp}$), and $\dot{n}_{\rm pa}>\dot{n}_{\rm pp}$ we should be safe from pair production. From these two considerations, the former is the most important.}
\label{tab:ppp}
\begin{tabular}{@{}lcccc}
\hline
fit			&$r0$ in $r_g$	&$\log$($n$)			&$\log$($\dot{n}_{\rm pa}\times r0/c$)			& $\log$($\dot{n}_{\rm pp}\times r0/c$)\\
\hline
{\sc smedge}	&20			&15.3				&12.9									& 15.0 \\
{\sc reflect}	&9.3			&16.1				&14.3									& 15.7\\
{\sc smedge2}	&6.4			&16.6				&15.1									& 15.8\\
{\sc 6kpc}		&4.9			&16.0				&14.9									& 15.1\\
{\sc eltemp}	&4.2			&17.1				&16.3									& 13.4\\ 
{\sc synch}	&3.6			&16.4				&14.5									& 15.2\\
\hline
\end{tabular}
\end{table}

\section{Discussion}\label{sec:grsdisc}
 
Clearly, and perhaps surprisingly, there are large differences between
datasets 1 and 2, both in the $\chi$ state of \grs.  While dataset 1 can be fit fairly well by the same model as for BHBs,
dataset 2 is marginal, thus they do not seem to represent a standard
class of spectral behaviour.   Interestingly, \cite{2003A&A...397..711R} also find
significant variety between the individual $\chi$ states they observed
in over 4 years.  While the classification done by
\cite{Bellonietal2000} is based on lightcurves and colour-colour
diagrams, the sub-indices 1-4 do not refer to a phenomenological classification scheme, but only denote their temporal
sequence. Hence there remains the possibility that the $\chi$ state
still encompasses multiple physical regimes that we may be probing
here via spectral fitting.  

Alternately, dataset 2 could be in some kind of transitional state,
despite being classified as a typical $\chi$ state.  In \grs\
complete state changes can happen on very short timescales in
comparison to the canonical BHs (seconds). Such rapid changes
generally occur only between states A and B, however.  While state C
episodes can also last for short periods of $<$100 seconds, the
current datasets involve only the longer (a day or more) intervals.
As indicated by the radio and NIR lightcurves, it is clearly possible that during the dataset 2 observations \grs\
was in a transitional state and therefore the model, which assumes a steady
outflow, is no longer applicable.  Either way, both of the
statistically convincing dataset 2 fits ({\sc synch} and {\sc eltemp})
have other problems in their physical interpretation, as discussed
above, and it is possible that the model simply cannot apply to
observations at such high luminosities (see Figure \ref{fig:HID}). 

For all of the fits presented, two parameters consistently stand out
by nature of their larger values when compared to applications of the
MNW05 model to BHBs: the location of initial particle acceleration in
the jets, $z_{\rm acc}$, and the energy partition parameter $k$.  As
mentioned before, we typically find $z_{\rm acc}$ 1-2 orders
of magnitude closer in towards the base of the jets in fits of other
BHBs in the HS.  Similarly, while the canonical BHBs often display
mild magnetic domination over the gas pressure, with $k\simeq10$, we
find at least an order of magnitude increase is needed to fit \grs\
plateau states. Finally, the overall powers required (as indicated by
the jet normalisation parameter $N_j$, see MNW05 for an explanation)
are strikingly larger than the maximum observed in other BHBs.
Canonical BHBs displaying these luminosity levels in $L_{\rm Edd}$
would long have shut down their jets, yet somehow \grs\ seems
to be operating on a different energy scale.

These three parameters may well be related.  Magnetic fields are by
now known to play a major role in accelerating and collimating
relativistic plasma outflows, and current simulations of jet formation
favour rather high values of $k$ (usually expressed as low values of
the plasma $\beta$, denoting the ratio to of the magnetic to gas pressure; e.g. \citealt{McKinney2006,Komissarovetal2007,McKinneyBlandford2009}).  It may be
that $k$ is positively correlated with the overall power input into
the jets.  Evidence for such a scaling has already been observed in
individual fits to broadband data of the LLAGN M81* over the course of a
yearlong campaign \citep[see, e.g.][Figure 22]{Markoffetal2008}.
The overall magnetic field strength and configuration will certainly
influence the formation of particle acceleration structures, however
the dependence on total power and other parameters has not yet been
thoroughly explored. %(but see Polko et al., in prep.).  

A higher level of magnetic domination can also account for why the
fits favour an electron temperature $T_e$ that is a factor of two or
more lower than the the bottom of the range ($\sim2-7\times10^{10}$ K)
found for the canonical microquasars.  A stronger magnetic field
relative to the radiating plasma at the base of the jets in our model
is synonymous with the same being true at the inner edge of the
accretion flow, as we assume they are directly related.  A relatively
stronger magnetic field would result in relatively more lepton
cooling in the inner regions, and consequently a lower equilibrium temperature.  

In contrast, the geometry of both the base of the jets and the cool
accretion disc parameters preferred in the fits is entirely consistent
with the range found in the canonical BHBs.  The radius of the base of
the jets is $3-20\;r_g$ exactly as seen in fits of both BHBs as well as
LLAGN.  Similarly, we obtain best fit values for the inner cool
accretion disc radius $r_{\rm in}\sim2.5-4.7 r_g$, which is consistent
with such a high accretion rate, as well as a spinning black hole as
suggested by \citet{2006ApJ...652..518M}.  What is not consistent, and
differs compared to other BHB fits, is that the jet base radius in
most fits is larger (sometimes only marginally) than $r_{\rm
  in}$.  Since in our model the jets should be launched by the inner,
hot accretion flow, this is not internally self-consistent and by such
criteria, only fits {\sc smedge2} and {\sc eltemp} would survive.  As
the MNW05 model does not yet self-consistently solve the jet launching
physics from a corona, which is still an unsolved problem, we have not made this a hard constraint.

Although in most BHBs such a small $r_{\rm in}$ would be indicative of
a soft state, this is not true for \grs. Due to the high
accretion rate, a geometry where the accretion disc reaches the
innermost stable orbit is almost expected. The fact that the {\sc
  6kpc} fit has an inner radius of less then 1 $r_g$ is most certainly
due to the oversimplifications mentioned above and similar inner radii
were also obtained for the canonical BHs (see e.g. MNW05). In fact,
such small radii appear to be a common occurrence, also using other
models: analysing 4 years of $\chi$ state observations
\citet{2003A&A...397..711R} obtained inner radii of far less then 1
r$_g$, from the normalisation of the disc contribution using the
\texttt{diskbb} model. Although this model is known to underestimate
the inner disc radius by a factor of 1.7-3, due to Doppler blurring
and gravitational redshift, the obtained radii are still
unrealistically low. The disc temperatures of 1-4 keV they find are
higher than ours (0.7-0.9 keV), but again the \texttt{diskbb} model is
known to produce factor 1.7 overestimates
\citep{2002ApJ...567.1102L}. Comparing the two plateau states to each
other, the required disc contribution is much higher in dataset 2. For
dataset 1, however, disc normalisations are high but comparable to
those for the canonical BHBs, suggesting that, as is true for the HS,
the classification of individual plateau states should not be based on
the luminosity.

In agreement with the findings of many other authors
(e.g. \citealt{2003A&A...400..553S,2003A&A...397..711R}) who have
studied the $\chi$ state, we find that the inverse Compton component
is always dominant over the direct accretion disc contribution in the
X-ray regime for our datasets. Although in our case the Comptonisation
is mainly due to SSC rather than purely from thermal accretion disc
seed photons. Only in one dataset 2 fit ({\sc synch}) are the X-rays not dominated by the
Comptonisation, but the validity of this fit may be questionable on
other grounds as discussed above.

As brought up as a potential issue by \citet{Maitraetal2009a} and
\citet*{MalzacBelmontFabian2009}, for high luminosities conditions at the base of the
jet may comprise high enough photon
densities that pair production can become important.  As these
processes are not yet implemented in the MNW05 model, we check the
pair production and annihilation rates ($\dot{n}_{pp}$ and
$\dot{n}_{pa}$ respectively) at the base of the jets using the methods
described in \citet{Maitraetal2009a}, and references therein. The
results are shown Table \ref{tab:ppp}. If the lepton number density $n$
is not larger than $\dot{n}_{pp} r_0/c$ and $\dot{n}_{pa} r_0/c$,
where $r_0/c$ is roughly the \emph{residence time} for the leptons in the jet
base region, we need $\dot{n}_{pp}<\dot{n}_{pa}$ to be able to neglect
pair production.

Although our $T_e\lesssim10^{10}$ is a factor of 5--6 lower than the limit
deemed problematic for GX339-4, the extreme luminosity of \grs\ means
that pair production could potentially be an issue for some of our fits.  The
source of potential pair production is a high-energy tail above
$\sim0.5$ MeV, due to the Comptonisation of thermal accretion disc
photons. The amount of flux in this tail is directly dependent on the
normalisation and temperature of the accretion disc component, which
itself is mainly constrained by the value of $N_{\rm H}$ and the
predicted flux in the tail region.  Higher values of $N_{\rm H}$
result in higher accretion disc fluxes to compensate for the increased
absorption in the soft X-ray band.

While we fixed the value for $N_{\rm H}$ in our X-ray fits to $4.7\times10^{22}$ cm$^{-2}$, an increasing number of works are
concluding that the column density is actually variable, ranging in
values from $N_{\rm H}\sim 2-16\times10^{22}$ cm$^{-2}$ and
potentially linked with an intrinsic warm absorber at the highest
values
\citep{Bellonietal2000,2002MNRAS.331..745K,2002ApJ...567.1102L,2006ApJ...646..385Y}.
Although generally the column density derived from X-ray measurements is higher than those obtained using IR methods, we find that the highest values of the $N_{\rm H}$ range
 are not
consistent with our model. Values $\gtrsim7$ cause too great a decrease below 3
keV, resulting in too much flux in the high-energy tail, due to enhanced disc Comptonization. 
We chose the above more ``average" value for $N_{\rm H}$, in order to be
consistent with prior X-ray fits of the various $\chi$ substates
(e.g. 2$\times10^{22}$ cm$^{-2}$; \citealt{Bellonietal1997b}) and
the assumptions of \citet{1996A&A...310..825C}. For dataset 2 the above 
value is likely too high.  Letting $N_{\rm H}$ free to vary (without
correcting the NIR dereddening accordingly), while maintaining a higher electron temperature of $\sim10^{10}$ K settles on the lower bound allowed of $2\times10^{22}$ cm$^{-2}$, with a significant improvement in
$\chi_{\rm red}^2$ (from $\sim$30 to $\sim25$), mainly due to the resulting reduction
in the Comptonised high-energy tail.  Therefore our results are highly dependent on the value of $N_{\rm H}$, and
that this value is likely different for the two datasets, although for
the reasons described above we have chosen to use a single value.
Similarly, the uncertainty in the exact value means
that too much pair production can be avoided in particular by lower
values of $N_{\rm H}$.  Lower values are favoured if we allow $N_{\rm
  H}$ to vary, and interestingly this is consistent with the
conclusion by \citet{2009Natur.458..481N} that the presence of an
absorbing disc wind at high accretion rates should be anticorrelated
with the presence of jets.

\section{Conclusions}
\label{sec:grsconcl}

Despite the fact that the MNW05 model was originally intended for
application to hard states in canonical BHBs only, it appears to well
approximate the plateau state in \grs. However it does not
produce convincing results in every instance. While some of the
parameter values obtained are quite extreme, the results for the
dataset 1 appear credible and consistent with what we have found in
canonical BHBs.  Dataset 2 however presents more difficulties, and a
more solid determination of the distance and absorption column would
go far to help us understand the difference between these two plateau
states.  

Clearly this work confirms that \grs\ is in a much more extreme
range of parameter space for an outflow-dominated model, requiring near- or 
super-Eddington accretion rates, maximal jet powers and high levels of
magnetic domination.  Although our results confirm previously noted
plateau state issues, such as the need for a variable $N_{\rm H}$,
our model for the first time incorporates the entire broadband and has
allowed the comparison between the jet producing plateau and hard
states.  While the baseline geometry seems similar, the plateau states
of \grs\ are not low-luminosity as with HS BHBs, and settle on a range
where the acceleration of particles occurs much further out in the
jets, which can be two orders of magnitude further out of
equipartition in the direction of magnetic domination.  While these
two effects may be linked, the model explored here cannot
self-consistently address this, but in another work we are exploring
the links between physical parameters and the location of particle
acceleration fronts (Polko et al., in prep.).  A slightly lower
electron temperature is also found compared to other BHBs, which can
be interpreted in the context of the higher cooling rates found at 
\grs's extreme luminosity.  

The main consequence of these differences is that the synchrotron
component from the outer jets no longer dominates the soft X-ray band,
although a non-negligible X-ray synchrotron flux of $\sim10$ \% the
inverse Compton flux below $\sim 50$ keV seems required for the
statistically favoured fits. In comparison, the MNW05 model applied to
canonical BHBs favours synchrotron emission dominating the flux at
least up to 10 keV.  Interestingly our results are thus qualitatively
similar to the results found from the blazar sequence
\citep[e.g.][]{Ghisellinietal1998}, where higher powers correspond to
a decrease in the frequency range where synchrotron power peaks.  It
is clear that time-dependent effects such as cooling breaks demand
further exploration, and we are currently working to implement them
into more complex models.

The remarkable differences between two individual plateau state
observations does raise questions about whether the $\chi$ substates
are distinct enough to classify all plateau characteristics.  While
both datasets explored here bear all the characteristics of the
plateau state, the fits with a single model show more variations in
free parameters than found even between different sources in the HS of
canonical BHBs.  Thus the current classification scheme based solely
on X-ray colours and timing properties may need to be expanded based on
broadband attributes.  

Our results support a conclusion that, although expressing quite
different properties than the HS in canonical BHBs, \grs\ plateau
states can still be described by the same broadband model with a
steady outflow tied in power to the accretion inflow.  However the BHB
model is clearly forced into very extreme ranges, which themselves provide 
some new clues about the relationship between accretion rate, jet
production and particle acceleration.  

Although \grs\ is one of the most extensively studied BHBs over the
last decades since its discovery, we are far from understanding this
source.  At some point in the future (assuming the source went into outburst when it was first discovered before the end of the century; \citealt*{DeeganCombetWynn2009}) \grs\ will
invariably retreat into quiescence, and should finally yield better
insight into the connection between its current accretion properties
and those at sub-Eddington rates.

\section{Acknowledgments}

The Green Bank Interferometer is a facility of the National Science Foundation operated by the NRAO in support of NASA High Energy Astrophysics programs. 

SM and DM acknowledge support from a Netherlands Organization for Scientific Research (NWO) Vidi and VC 
Fellowship, respectively

The research leading to these results has received funding from the 
European Community's Seventh Framework Programme (FP7/2007-2013) under 
grant agreement number ITN 215212 ``Black Hole Universe"

AJCT acknowledges support from the Spanish Ministry program AYA 2009-14000-C03-01

\bibliographystyle{mn}
\bibliography{citation}

\end{document}